\RequirePackage{fix-cm}
\documentclass[twocolumn,epjc3]{svjour3}  
\smartqed  
\RequirePackage{graphicx}
\journalname{Eur. Phys. J. C}
\usepackage{cite} 
\usepackage{amsmath,amssymb}
\usepackage{color}
\usepackage{xcolor}
\usepackage[caption=false]{subfig}
\usepackage{float}
\usepackage{placeins}
\usepackage{booktabs}
\usepackage{array}
\usepackage{hyperref}

\begin{document}
\sloppy
\title{Radial pulsations of gravitationally decoupled anisotropic self-bound stars with a generalized linear equation of state}

\titlerunning{Radial pulsations of anisotropic self-bound stars}        

\author{D. Santana
\thanksref{e3,addr3} 
        \and
        C. Usca\thanksref{e4,addr3} 
        \and
        J. Andrade\thanksref{e5,addr3}
        \and
        J. O. Cedeño\thanksref{e6,addr3,addr4}
}
\thankstext{e3}{e-mail: 
\href{mailto:diego.santana@espoch.edu.ec}{\nolinkurl{diego.santana@espoch.edu.ec}}}
\thankstext{e4}{e-mail: 
\href{mailto:icrisus25@gmail.com}{\nolinkurl{icrisus25@gmail.com}}}
\thankstext{e5}{e-mail: 
\href{mailto:julio.andrade@espoch.edu.ec}{\nolinkurl{julio.andrade@espoch.edu.ec}}}
\thankstext{e6}{e-mail: 
\href{mailto:jorge.cedeno@czetesis.com}{\nolinkurl{jorge.cedeno@czetesis.com}}}

\institute{Escuela Superior Politécnica de Chimborazo, (ESPOCH), Riobamba 060155, Ecuador\hypertarget{addr3}{}\label{addr3}
\and
 Centro de Investigación y Desarrollo Científico Zétesis, Ecuador\hypertarget{addr4}{}\label{addr4}} 

\date{Received: date / Accepted: date}

\maketitle

\begin{abstract}
We construct anisotropic self-bound stellar configurations through gravitational decoupling in the minimal geometric deformation framework and investigate their stability against linear radial perturbations. Retaining the Tolman IV temporal potential, we impose the generalized linear equation of state $P_r=\eta(\rho-\rho_s)$ on the complete matter distribution. Central regularity and Schwarzschild matching determine the radial geometry and relate the surface density to the compactness and stellar radius. Closed-form solutions are obtained for $\eta=1,1/2,1/3,1/4$, including the MIT bag radial relation when $\eta=1/3$ and $\rho_s=4B$. Since the equilibrium anisotropy does not determine its dynamical response, we introduce two material prescriptions that preserve either the anisotropy of each fluid layer or its ratio to $\rho+P_r$. Including metric perturbations, we compute the fundamental mode and three overtones for twelve equilibrium configurations under both prescriptions. All 24 background response combinations have positive fundamental squared frequencies, establishing linear radial stability within the adopted closures. The fundamental eigenvalue decreases with compactness within each sampled family, while changing the anisotropic response produces shifts whose direction depends on the background. Tangential tensions in several configurations and the failed tangential profile diagnostic of the maximally stiff family limit their physical interpretation. These results provide analytically tractable self-bound backgrounds and demonstrate that identical equilibrium configurations can support distinct radial spectra depending on their anisotropic material response.
\end{abstract}


\section{Introduction}\label{intro}

Compact stars constitute unique astrophysical laboratories for investigating the
behaviour of matter and gravity under conditions that cannot be reproduced in
terrestrial experiments \cite{camenzind2007compact}. Their interiors may reach densities several times higher than the nuclear saturation density ($\sim 2.7\times 10^{17}\,\mathrm{kg\,m^{-3}}$ in mass-density units \cite{huth2022constraining}), where the microscopic composition of matter remains uncertain and a wide variety of phases may arise, including hyperonic matter, meson condensates, deconfined quarks, color-superconducting phases, and other strongly interacting states
\cite{glendenning2000compact,haensel2007neutron,shapiro1983implications,lattimer2016equation,weber2005strange}. The macroscopic properties of these
objects, such as their masses, radii, compactness, tidal response, and oscillation
spectra, depend strongly on the equation of state (EoS) governing the stellar
interior. Consequently, constructing relativistic stellar models that connect
the microscopic properties of dense matter with potentially observable
quantities remains one of the central problems of modern gravitational physics
and high-energy astrophysics.

For a fixed EoS, the central density typically increases with gravitational
mass along a connected stable branch of nonrotating stellar configurations,
making massive pulsars particularly valuable for probing matter under extreme
conditions. Among the landmark measurements is that of PSR J1614$-$2230,
whose mass was initially determined to be
\(1.97\pm0.04\,M_{\odot}\) using Shapiro delay
\cite{demorest2010two} and subsequently refined to
\(1.928\pm0.017\,M_{\odot}\) in the NANOGrav nine-year analysis
\cite{fonseca2016nanograv}. Another important measurement is that of
PSR J0348$+$0432, with \(2.01\pm0.04\,M_{\odot}\)
\cite{antoniadis2013massive}. The existence of compact stars with masses close
to or above two solar masses places strong constraints on the EoS, which must
provide sufficient pressure at supranuclear densities to prevent gravitational
collapse \cite{cromartie2020relativistic}.

While the maximum mass of a compact star is mainly determined by the behaviour
of matter in its high-density core, the radii of predominantly nucleonic neutron
stars with canonical masses of approximately \(1.3\)--\(1.5\,M_{\odot}\)
are particularly sensitive to the pressure around and above nuclear saturation
density, and hence to the density dependence of the nuclear symmetry energy
\cite{lattimer2016equation}. Precise radius measurements are therefore essential
for constraining dense matter, although their interpretation depends on the
observational method and on assumptions concerning thermal emission, stellar
atmospheres, source distance, geometry, and X-ray pulse profiles. The NICER
mission \cite{gendreau2016nicer} has provided mass--radius constraints through
pulse-profile modelling, including analyses of PSR J0740$+$6620
\cite{miller2021radius,riley2021radius}. The scientific cases for X-ray mission
concepts such as Athena+ and LOFT have also emphasized the potential of
high-throughput spectroscopy and timing for studying compact objects
\cite{nandra2013athena,feroci2012loft}. Complementary information has been
obtained from the multimessenger observation of GW170817 and its associated
kilonova \cite{abbott2017multimessenger}, while analysis of the gravitational-wave
signal has constrained tidal deformabilities and the dense-matter EoS
\cite{abbott2018gw170817}. Nevertheless, such macroscopic constraints do not
by themselves provide a unique identification of the stellar composition:
hybrid stars can have mass--radius relations resembling those of hadronic
stars \cite{alford2005masquerade}, and self-bound quark-star models have also
been constructed that satisfy mass, radius, and tidal constraints
\cite{lourenco2021tidal}. This motivates the investigation of complementary
diagnostics, including oscillation spectra and dynamical stability
\cite{vazquezflores2010radial,arbanil2016anisotropic,arbanil2023tidal}.

Strange quark stars constitute a theoretically motivated, although still
hypothetical, class of compact objects
\cite{alcock1986strange,haensel1986strange,weber2005strange}.
Their possible existence is based on the Bodmer--Witten hypothesis, according
to which bulk matter composed of up, down, and strange quarks may be
absolutely stable and represent the true ground state of strongly interacting
matter
\cite{bodmer1971collapsed,witten1984cosmic,farhi1984strange}.
Unlike ordinary neutron stars, which are bound primarily by gravity, strange
quark stars would be self-bound by strong interactions and could retain a
finite density at zero pressure, leading to distinctive internal structures
and mass--radius relations
\cite{alcock1986strange,haensel1986strange,glendenning2000compact,weber2005strange}. Historically, radius estimates for sources such as
SAX J1808.4$-$3658 and RX J1856.5$-$3754 motivated interpretations in terms
of strange or quark stars \cite{li1999sax,drake2002quark}; these proposals
should not be regarded as conclusive identifications. Deconfined quark
matter may instead be confined to the cores of hybrid stars
\cite{alford2013generic}, whose equilibrium mass--radius relations can closely
resemble those of purely hadronic stars \cite{alford2005masquerade}.
Conversion to quark matter has also been proposed as an energy source in
models of superluminous supernovae \cite{leahy2008supernova} and gamma-ray
bursts \cite{paczynski2005gamma}. Other proposed signatures concern
gravitational waves from compact-star mergers \cite{bauswein2010mergers},
thermal evolution \cite{page2002thermal}, and stellar oscillations
\cite{vazquezflores2010radial,arbanil2016anisotropic,arbanil2023tidal}.

The macroscopic description of self-bound quark matter commonly employs the MIT
bag model or generalized linear and constant-speed-of-sound equations of state
\cite{chodos1974new,farhi1984strange,zdunik2000strange,alford2013generic,alford2015quark}. Although phenomenological, these models
capture essential features of dense quark matter, including its stiffness and
finite surface density, while allowing different physical regimes to be explored
within a unified framework. A generalized linear EoS is particularly useful
because it can reproduce the simplest MIT bag relation, approach the causal
stiffness limit, or interpolate between softer and stiffer self-bound phases.
It therefore provides a flexible basis for studying how the properties of dense
matter influence the mass, radius, compactness, oscillation spectrum, and
stability of strange quark stars without restricting the analysis to a single
microscopic model.

Most relativistic stellar models treat matter as a perfect fluid, with the same
pressure in every direction. This is a useful starting point, but it may leave
out stresses associated with the internal structure of strange quark stars.
Strong magnetic fields and crystalline phases, for example, can introduce
preferred directions and allow the radial and tangential pressures to differ.
Phase boundaries, density gradients, and the coexistence of different forms of
matter may also contribute, depending on their microscopic properties. Pressure
anisotropy thus has physical motivations, although its presence and magnitude
must be assessed for the particular model rather than inferred from high density
alone
\cite{bowers1974anisotropic,ruderman1972pulsars,canuto1977quantizing,herrera1997local,ivanov2002static,doneva2012nonradial,deb2021anisotropic}. A related issue arises during gravitational collapse:
studies of spherically symmetric dissipative fluids show that dissipative
fluxes, density inhomogeneities, and shear can disturb an initially isotropic
pressure distribution \cite{herrera2020stability}. These results provide a
further reason to examine anisotropic models, but they do not establish a
universal anisotropic end state for compact stars..

When radial and tangential pressures differ, their difference contributes to
the force balance inside the star. A tangential pressure greater than the radial
pressure provides an outward contribution, whereas the reverse produces an
inward one. This changes the equilibrium structure and can affect the mass a
star can support, its radius and compactness, and consequently its surface
redshift and tidal response. Anisotropy also influences how the star responds
to perturbations: it can change the oscillation frequencies and the point at
which radial instability sets in. The size and direction of these changes
depend on the assumed anisotropy and, for oscillations and stability, on how
the stresses respond when the fluid is perturbed
\cite{dev2002anisotropic,dev2003anisotropic,horvat2011radial,arbanil2016anisotropic,arbanil2023tidal}.

Allowing the radial and tangential pressures to differ introduces an extra
unknown into the stellar structure equations. Einstein's equations must
therefore be supplemented by a relation that specifies the anisotropy or
otherwise connects the matter variables.
Several analytical techniques have been developed for this purpose, including
prescribed metric potentials, generating functions, symmetry conditions,
complexity constraints, and phenomenological relations among the thermodynamic
variables \cite{tolman1939static,buchdahl1959general,lake2003generating,herrera2018new,maurya2019anisotropic,ovalle2008searching}. Among these approaches, gravitational decoupling (GD) \cite{ovalle2008searching} has emerged as a systematic framework for introducing additional matter contributions into known stellar solutions.

From a technical point of view, the anisotropic configurations considered in
this work are constructed through GD in its minimal geometric deformation formulation (MGD) \cite{ovalle2017decoupling}. This framework separates the total matter distribution into a known seed source and an additional sector responsible for deforming the original geometry, thereby allowing anisotropic extensions of well-behaved relativistic solutions to be obtained in a controlled manner \cite{ovalle2008searching,ovalle2017decoupling,ovalle2018anisotropic,ovalle2019extended}. The method has been widely applied to compact objects
containing perfect fluids, polytropic matter, electromagnetic fields, scalar
sources, and other effective matter distributions
\cite{gabbanelli2018anisotropic,contreras2018minimal,telloortiz2020anisotropic,leon2021gravitational,santana2022integration,andrade2024gravitational,naseer2026polytropes,andrade2025implications,andrade2021stellar,siza2024anisotropic}. Nevertheless, compliance with regularity, energy, causality, and matching conditions does not guarantee dynamical stability, since an apparently acceptable equilibrium configuration may become unstable under small perturbations.

For this reason, the present study complements the gravitational decoupling
construction with an analysis of radial pulsations. The relativistic theory
developed by Chandrasekhar describes these oscillations through a discrete
spectrum of normal modes, with the fundamental frequency determining the onset
of instability
\cite{chandrasekhar1964dynamical,chandrasekhar1964stability,gondek1997radial,kokkotas1999quasi,luz2024adiabatic}. Radial oscillations of self-bound quark stars have previously been studied for the MIT bag model,
color-superconducting matter, perturbative quantum chromodynamics, and
multicomponent configurations
\cite{vazquezflores2010radial,panotopoulos2017radial,jimenez2019radial,bora2021radial}. In anisotropic systems, however, the
difference between radial and tangential stresses modifies both the equilibrium
structure and the restoring forces, potentially shifting the maximum stable
mass and the point at which the fundamental mode vanishes
\cite{dev2003anisotropic,horvat2011radial,arbanil2016anisotropic,arbanil2023tidal,mohanty2024unstable}. Accordingly, the
present work investigates self-bound anisotropic stars governed by a generalized
linear EoS, with the aim of determining how gravitationally induced anisotropy
affects their equilibrium properties, radial oscillation spectrum, and
dynamically stable domain.

The paper is organized as follows. Section~\ref{section-field-equations}
introduces the relativistic stellar framework and the effective matter
variables. Section~\ref{section-eos} presents the generalized self-bound EoS,
while Section~\ref{section-mgd} outlines the gravitational decoupling procedure
used to construct the anisotropic configurations.
Section~\ref{section-physical-analysis} examines their physical properties and
matching conditions. The radial pulsation formalism and stability analysis are
developed in Section~\ref{section-radial-pulsations}. The main conclusions are summarized in Section~\ref{section-conclusions}.

\section{Einstein's field equations for an anisotropic relativistic fluid}
\label{section-field-equations}

We consider a static and spherically symmetric self-gravitating configuration
described by Einstein's field equations (EFEs),
\begin{equation}
    G^{\mu}_{\ \nu}
    =
    \frac{8\pi G}{c^{4}}\,T^{\mu}_{\ \nu},
    \label{eq:EFE}
\end{equation}
where \(T^{\mu}_{\ \nu}\neq 0\) and represents the energy--momentum tensor associated with
the matter distribution. Throughout the remainder of this work, geometrized
units \(G=c=1\) are adopted. The interior spacetime is represented by the line
element
\begin{equation}
    ds^{2}
    =
    e^{\nu(r)}dt^{2}
    -
    e^{\lambda(r)}dr^{2}
    -
    r^{2}d\Omega^{2},
    \label{eq:Sch}
\end{equation}
where
\begin{equation}
    d\Omega^{2}
    =
    d\theta^{2}
    +
    \sin^{2}\theta\,d\phi^{2},
\end{equation}
and the metric potentials \(\nu(r)\) and \(\lambda(r)\) depend only on the
radial coordinate.

The matter sector is modelled as an anisotropic fluid whose energy--momentum
tensor is written as
\begin{equation}
    T_{\mu\nu}
    =
    \left(\rho+P_{\perp}\right)u_{\mu}u_{\nu}
    -
    P_{\perp}g_{\mu\nu}
    +
    \left(P_{r}-P_{\perp}\right)s_{\mu}s_{\nu},
    \label{eq:TensorEM}
\end{equation}
where \(\rho\) denotes the energy density, while \(P_r\) and \(P_{\perp}\)
represent the radial and tangential pressures, respectively. For a comoving
observer, the four-velocity and the unit radial vector are given by
\begin{equation}
    u^{\mu}
    =
    \left(e^{-\nu/2},0,0,0\right),
    \label{eq:velocidad-fluido}
\end{equation}
and
\begin{equation}
    s^{\mu}
    =
    \left(0,e^{-\lambda/2},0,0\right),
    \label{eq:su}
\end{equation}
which satisfy
\begin{equation}
    u^{\mu}u_{\mu}=1,
    \qquad
    s^{\mu}s_{\mu}=-1,
    \qquad
    u^{\mu}s_{\mu}=0.
\end{equation}

Substituting Eqs.~\eqref{eq:Sch} and \eqref{eq:TensorEM} into
Eq.~\eqref{eq:EFE} yields the independent field equations
\begin{align}
    8\pi\rho
    &=
    \frac{1-e^{-\lambda}}{r^{2}}
    +
    \frac{e^{-\lambda}\lambda'}{r},
    \label{eq:metrica-EFE-1}
    \\[2mm]
    8\pi P_{r}
    &=
    -\frac{1-e^{-\lambda}}{r^{2}}
    +
    \frac{e^{-\lambda}\nu'}{r},
    \label{eq:metrica-EFE-2}
    \\[2mm]
    8\pi P_{\perp}
    &=
    \frac{e^{-\lambda}}{4}
    \left[
        2\nu''
        +
        \nu'^{2}
        -
        \lambda'\nu'
        +
        \frac{2(\nu'-\lambda')}{r}
    \right],
    \label{eq:metrica-EFE-3}
\end{align}
where primes denote derivatives with respect to \(r\).

In an anisotropic configuration, the radial pressure differs from the angular
components,
\begin{equation}
    P_{r}(r)
    \neq
    P_{\theta}(r)
    =
    P_{\phi}(r)
    \equiv
    P_{\perp}(r),
\end{equation}
where the equality between the two angular pressures follows directly from
spherical symmetry \cite{dev2002anisotropic}. It is therefore convenient to
introduce the anisotropy factor
\begin{equation}
    \Delta(r)
    =
    P_{\perp}(r)-P_{r}(r).
    \label{eq:anisotropy-factor}
\end{equation}
The isotropic limit is recovered when \(\Delta=0\).

The conservation of the energy--momentum tensor follows from the contracted
Bianchi identities,
\begin{equation}
    \nabla_{\mu}T^{\mu\nu}=0,
    \label{eq:leyconservacion}
\end{equation}
and leads to the generalized Tolman--Oppenheimer--Volkoff equation
\begin{equation}
    P_{r}'
    =
    -
    \frac{\nu'}{2}
    \left(\rho+P_{r}\right)
    +
    \frac{2}{r}
    \left(P_{\perp}-P_{r}\right).
    \label{eq:pr'}
\end{equation}
This expression describes the local equilibrium among the pressure gradient,
the gravitational contribution, and the force generated by pressure
anisotropy.

Introducing the mass function \(m(r)\) through
\begin{equation}
    e^{-\lambda(r)}
    =
    1-\frac{2m(r)}{r},
    \label{eq:masa1}
\end{equation}
the first field equation gives
\begin{equation}
    \frac{dm}{dr}
    =
    4\pi r^{2}\rho.
    \label{eq:mass-derivative}
\end{equation}
The temporal metric potential can then be expressed as
\begin{equation}
    \nu'
    =
    2
    \frac{
        m+4\pi r^{3}P_{r}
    }{
        r(r-2m)
    }.
    \label{eq:v'}
\end{equation}
Consequently, Eq.~\eqref{eq:pr'} becomes
\begin{equation}
    P_{r}'
    =
    -
    \frac{
        m+4\pi r^{3}P_{r}
    }{
        r(r-2m)
    }
    \left(\rho+P_{r}\right)
    +
    \frac{2\Delta}{r}.
    \label{eq:pr'final}
\end{equation}
Equation~\eqref{eq:pr'final}, together with
Eq.~\eqref{eq:mass-derivative}, governs the hydrostatic equilibrium of the
anisotropic configuration.

At the center of the star, regularity requires
\begin{equation}
    m(0)=0,
    \qquad
    \rho(0)=\rho_{c},
    \qquad
    P_{r}(0)=P_{\perp}(0),
    \label{eq:central-conditions}
\end{equation}
where \(\rho_c\) denotes the central energy density. The last condition implies
that the anisotropy must vanish at the center.

The interior geometry is matched at the stellar surface
\(r=R\equiv r_{\Sigma}\) to the exterior Schwarzschild spacetime \cite{darmois1927equations,israel1966singular},
\begin{equation}
    ds^{2}
    =
    \left(1-\frac{2M}{r}\right)dt^{2}
    -
    \left(1-\frac{2M}{r}\right)^{-1}dr^{2}
    -
    r^{2}d\Omega^{2},
    \label{eq:metrica-distribucion}
\end{equation}
where \(M=m(R)\) is the total gravitational mass. In the absence of a thin
shell at the boundary, continuity of the first and second fundamental forms
leads to
\begin{align}
    e^{\nu(R)}
    &=
    1-\frac{2M}{R},
    \label{eq:enusigma}
    \\
    e^{-\lambda(R)}
    &=
    1-\frac{2M}{R},
    \label{eq:lambdasigma}
    \\
    P_{r}(R)
    &=
    0.
    \label{eq:prsigma}
\end{align}
The condition \(P_r(R)=0\) determines the stellar radius during the numerical
integration, whereas Eq.~\eqref{eq:mass-derivative} gives \(m(R)=M\). Notice
that the vanishing of the radial pressure does not necessarily imply a
vanishing surface density. In particular, self-bound compact stars may satisfy
\begin{equation}
    \rho(R)=\rho_s\neq0,
\end{equation}
where \(\rho_s\) is the finite energy density at zero pressure.

The field equations presented above involve the metric potentials and the
thermodynamic variables \(\rho\), \(P_r\), and \(P_{\perp}\), and therefore
require additional physical relations to form a closed system. In the next
section, this closure is provided by a generalized linear equation of state
appropriate for self-bound compact matter. Later, within the gravitational
decoupling framework, these quantities will be interpreted as the corresponding
effective density and pressures of the complete anisotropic configuration.


\section{Generalized linear equation of state for self-bound matter}
\label{section-eos}

The equation of state constitutes the fundamental link between the microscopic
properties of dense matter and the macroscopic structure of a compact star. In
the case of strange quark matter, its precise form remains uncertain because the
densities reached in stellar interiors lie in a regime where nonperturbative
strong-interaction effects, the finite strange-quark mass, quark pairing, and
possible color-superconducting phases may become relevant
\cite{farhi1984strange,glendenning2000compact,weber2005strange,lugones2002cfl}. Consequently, phenomenological parametrizations that retain the
essential thermodynamic features of self-bound matter provide a useful
alternative to model-dependent microscopic descriptions.

A characteristic property of self-bound matter is that its pressure may vanish
at a finite energy density. This behaviour distinguishes strange quark stars
from conventional gravitationally bound neutron stars and follows naturally
from bag-like descriptions of deconfined quark matter
\cite{witten1984cosmic,farhi1984strange,alcock1986strange,haensel1986strange}. Moreover, detailed MIT bag calculations including a finite
strange-quark mass and strong-interaction corrections can be accurately
represented, over the range of densities relevant to compact stars, by a linear
relation between pressure and energy density
\cite{zdunik2000strange,harko2002maximum}. Similar constant-speed-of-sound
parametrizations have also been extensively employed in model-independent
studies of quark and hybrid stellar matter
\cite{alford2013generic,alford2015quark}.

Motivated by these considerations, we adopt the generalized linear equation of
state
\begin{equation}
P_{r}
=
\eta\left(\rho-\rho_{s}\right),
\qquad
0<\eta\leq1,
\qquad
\rho\geq\rho_{s},
\label{eq:generalized-linear-eos}
\end{equation}
where $P_r$ is the radial pressure, and $\rho$ denotes the energy density, 
$\eta$ is a dimensionless stiffness parameter, and $\rho_s$ represents the
energy density at zero pressure. In the present work, $\rho$ must be
understood as an energy density rather than a rest-mass density. Equation
\eqref{eq:generalized-linear-eos} may equivalently be written as
\begin{equation}
\rho
=
\rho_s+\frac{P_r}{\eta}.
\label{eq:inverse-generalized-eos}
\end{equation}
This form belongs to the constant-speed-of-sound class of equations of state and
separates two physically different effects: $\eta$ controls the response of
the pressure to compression, whereas $\rho_s$ determines the density scale at
which the pressure vanishes
\cite{zdunik2000strange,alford2013generic,alford2015quark}.

For a stellar configuration with central density $\rho_c$, the central radial
pressure is
\begin{equation}
P_c
=
\eta\left(\rho_c-\rho_s\right),
\qquad
\rho_c>\rho_s.
\label{eq:central-pressure-linear-eos}
\end{equation}
At the stellar surface $r=R$, the condition $P_r(R)=0$ leads directly to
\begin{equation}
\rho(R)=\rho_s.
\label{eq:surface-density-linear-eos}
\end{equation}
Therefore, configurations with $\rho_s>0$ possess a sharp surface with a
finite energy density and are self-bound. By contrast, when $\rho_s=0$, the
density also vanishes at the zero-pressure surface, and the corresponding
relation does not describe self-bound matter in the strict sense
\cite{alcock1986strange,haensel1986strange,weber2005strange}.

The radial speed of sound follows immediately from
Eq.~\eqref{eq:generalized-linear-eos},
\begin{equation}
v_{s,r}^{2}
=
\frac{dP_r}{d\rho}
=
\eta.
\label{eq:radial-sound-speed-linear-eos}
\end{equation}
In geometrized units, causality requires $v_{s,r}^{2}\leq1$, while local
mechanical stability requires $v_{s,r}^{2}>0$. These requirements justify the
interval
\begin{equation}
0<\eta\leq1.
\label{eq:eta-domain}
\end{equation}
The upper value corresponds to propagation at the speed of light and therefore
represents the maximally stiff causal limit rather than a generic microscopic
prediction
\cite{zeldovich1962ultrahigh,rhoades1974maximum,lattimer2016equation}. The frequently used value $\eta=1/3$, on the other
hand, is associated with an ultrarelativistic gas of noninteracting massless
particles and should not be interpreted as a universal upper bound for dense
matter.

For $\rho_s\geq0$ and $0<\eta\leq1$, the radial pressure automatically
satisfies
\begin{equation}
P_r\geq0,
\qquad
\rho-P_r
=
(1-\eta)\rho+\eta\rho_s
\geq0,
\label{eq:radial-dominant-energy-eos}
\end{equation}
throughout the stellar interior. Hence, the radial sector is compatible with
the inequalities $0\leq P_r\leq\rho$. Nevertheless, the corresponding
conditions for the tangential pressure are not fixed by the radial equation of
state and must be verified after constructing the complete anisotropic
solution.

Another quantity of particular relevance for the radial-pulsation analysis is
the relativistic adiabatic index,
\begin{equation}
\Gamma_r
=
\frac{\rho+P_r}{P_r}
\frac{dP_r}{d\rho}.
\label{eq:relativistic-adiabatic-index}
\end{equation}
Using Eq.~\eqref{eq:generalized-linear-eos}, it becomes
\begin{equation}
\Gamma_r
=
\eta\frac{\rho+P_r}{P_r}
=
1+\eta+\frac{\rho_s}{\rho-\rho_s}.
\label{eq:adiabatic-index-linear-eos}
\end{equation}
Although $\Gamma_r$ diverges formally when $P_r\rightarrow0$ at the surface
of a self-bound star, the combination that enters the pulsation equations
remains finite,
\begin{equation}
\Gamma_r P_r
=
\eta\left(\rho+P_r\right).
\label{eq:gamma-pressure-product}
\end{equation}
This property must be taken into account when imposing the surface boundary
condition in the radial-stability problem
\cite{chandrasekhar1964dynamical,gondek1997radial,vazquezflores2010radial,bora2021radial}.

The generalized relation \eqref{eq:generalized-linear-eos} contains several
physically relevant limiting cases, which are discussed below.

\subsection{MIT bag-model limit.}

In the simplest MIT bag description, cold quark matter is composed of
noninteracting and massless up, down, and strange quarks confined by a vacuum
energy density $B$ \cite{chodos1974new,farhi1984strange}. For the quark contribution,
\begin{equation}
\rho_q=3P_q,
\label{eq:massless-quark-relation}
\end{equation}
while the bag vacuum modifies the total energy density and pressure according
to
\begin{equation}
\rho=\rho_q+B,
\qquad
P_r=P_q-B.
\label{eq:bag-energy-pressure}
\end{equation}

The opposite signs of the vacuum contribution follow from its thermodynamic interpretation. The bag constant $B>0$ represents the excess vacuum energy density inside the bag relative to the exterior vacuum. For a volume $V$, the corresponding energy is $E_{\mathrm{vac}}=BV$. Since $B$ is constant, its pressure contribution is

\begin{equation}
P_{\mathrm{vac}} = -\frac{\partial E_{\mathrm{vac}}}{\partial V}
=
-B.
\label{eq:bag-vacuum-pressure}
\end{equation}

Thus, the bag vacuum increases the total energy density while reducing the pressure, reflecting the energy cost of expanding the region occupied by deconfined quark matter. Both contributions are dimensionally consistent, since pressure and energy density have the same physical dimensions.

Eliminating the quark variables gives
\begin{equation}
P_r
=
\frac{1}{3}\left(\rho-4B\right).
\label{eq:mit-bag-eos}
\end{equation}
Therefore, the simplest MIT bag radial relation is recovered from the generalized linear equation
of state by choosing
\begin{equation}
\eta=\frac{1}{3},
\qquad
\rho_s=4B.
\label{eq:mit-parameter-identification}
\end{equation}
The finite value $\rho(R)=4B$ reflects the self-bound character of the radial relation. In an anisotropic configuration, this identification does not by itself recover the complete stress tensor of an isotropic MIT bag fluid; the tangential stress requires a separate prescription. Finite quark masses, perturbative strong-interaction corrections,
and pairing effects modify this simple relation, but their resulting equations
of state can frequently be approximated by the same linear form with effective
values of $\eta$ and $\rho_s$
\cite{farhi1984strange,zdunik2000strange,lugones2002cfl,alford2015quark}.

\subsection{Conformal ultrarelativistic limit.}

For $\eta=\frac{1}{3}$ and $\rho_s=0$, the EoS becomes
\begin{equation}
P_r=\frac{\rho}{3}.
\label{eq:conformal-linear-eos}
\end{equation}
This reproduces the radial pressure--energy-density relation of an ideal conformal gas of massless particles. For the complete anisotropic fluid, however, $T^\mu{}_{\mu}=\rho-P_r-2P_\perp=-2\Delta$ when $P_r=\rho/3$, so a vanishing trace additionally requires $\Delta=0$. A relation between radial pressure and density is not enough to establish
whether the full matter sector is conformal, since the tangential stresses
also enter the trace of the energy--momentum tensor. The absence of a vacuum
offset has a separate consequence: the density vanishes when the radial
pressure reaches zero. Thus, although
Eqs.~\eqref{eq:mit-bag-eos} and \eqref{eq:conformal-linear-eos} give the same
radial sound speed, they describe different surface behaviour. The MIT bag
relation allows a finite density at zero pressure, as required for self-bound
matter, whereas the relation without an offset does not.

\subsection{Zel'dovich stiff-fluid limit.}

The choice $\eta=1$ and $\rho_s=0$, reduces Eq.~\eqref{eq:generalized-linear-eos} to the Zel'dovich equation of
state,
\begin{equation}
P_r=\rho.
\label{eq:zeldovich-eos}
\end{equation}
This relation describes the most rigid fluid compatible with relativistic
causality, since $dP_r/d\rho=1$
\cite{zeldovich1962ultrahigh,rhoades1974maximum}. It is useful as a limiting
case for estimating the largest pressure support allowed by causality.
Nevertheless, because the density vanishes when the pressure vanishes, the
unshifted Zel'dovich relation is not a self-bound equation of state.

\subsection{Causal self-bound limit.}

A self-bound generalization of the maximally stiff case is obtained by setting
\begin{equation}
\eta=1,
\qquad
\rho_s>0,
\end{equation}
which yields
\begin{equation}
P_r=\rho-\rho_s.
\label{eq:shifted-zeldovich-eos}
\end{equation}
This equation preserves the causal-limit sound speed while allowing the
pressure to vanish at the finite density $\rho_s$. It may therefore be viewed
as a shifted Zel'dovich relation or as the maximally stiff member of the
self-bound family. The distinction between
Eqs.~\eqref{eq:zeldovich-eos} and \eqref{eq:shifted-zeldovich-eos} will be
important when comparing gravitationally bound and self-bound configurations.

More generally, the choices
\begin{equation}
0<\eta<1,
\qquad
\rho_s>0,
\label{eq:general-self-bound-domain}
\end{equation}
describe a continuous family of causal self-bound equations of state. By
varying $\eta$, one may investigate how the stiffness of dense matter affects
the stellar mass, radius, compactness, and radial oscillation spectrum, while
$\rho_s$ independently controls the density scale associated with the stellar
surface. This separation is one of the principal advantages of the generalized
linear parametrization.

It should be emphasized that Eq.~\eqref{eq:generalized-linear-eos} constrains
only the radial thermodynamic sector. It does not prescribe the tangential
pressure or the anisotropy factor. In the present model, these quantities will
be generated by the additional gravitational source introduced through the
minimal geometric deformation method. Thus, in the following section, the
generalized linear equation of state will be imposed on the effective radial
pressure and energy density of the complete gravitationally decoupled
configuration, providing the closure relation required to determine the
geometric deformation.


\section{Gravitational decoupling through minimal geometric deformation}
\label{section-mgd}

In this section, we introduce the GD framework through the MGD approach
\cite{ovalle2008searching,ovalle2017decoupling,ovalle2018anisotropic,ovalle2019extended}. In this method the energy--momentum tensor that appears in EFEs decomposes into a known seed gravitational source and an additional matter sector,
\begin{equation}
    T_{\mu\nu} = \widehat{T}_{\mu\nu} + \Theta_{\mu\nu},
    \label{eq:T-munu-mgd}
\end{equation}
where \(\widehat{T}_{\mu\nu}\) describes the seed configuration and
\(\Theta_{\mu\nu}\) represents the effective source responsible for the
deformation of the original geometry. In the present formulation, the usual
coupling parameter is absorbed into the additional source and into the
geometric deformation. Thus, \(\Theta_{\mu\nu}\) and the deformation function
introduced below represent the complete physical contribution of the
decoupling sector.

The seed source is usually assumed to correspond to an isotropic perfect fluid,
\begin{equation}
    \widehat{T}^{\mu}_{\ \nu}
    =
    \mathrm{diag}
    \left(
        \widehat{\rho},
        -\widehat{P},
        -\widehat{P},
        -\widehat{P}
    \right),
    \label{eq:seed-energy-momentum}
\end{equation}
where \(\widehat{\rho}\) and \(\widehat{P}\) denote the seed density and
pressure, respectively. The effective variables associated with the additional
source are defined as
\begin{equation}
    \rho_{\Theta}
    =
    \Theta^{0}_{\ 0},
    \qquad
    P_r^{\Theta}
    =
    -\Theta^{1}_{\ 1},
    \qquad
    P_{\perp}^{\Theta}
    =
    -\Theta^{2}_{\ 2}.
    \label{eq:theta-effective-variables}
\end{equation}
Consequently, the density and pressures of the complete configuration are
\begin{align}
    \rho
    &=
    \widehat{\rho}
    +
    \rho_{\Theta},
    \label{eq:rho-effective-mgd}
    \\
    P_r
    &=
    \widehat{P}
    +
    P_r^{\Theta},
    \label{eq:pr-effective-mgd}
    \\
    P_{\perp}
    &=
    \widehat{P}
    +
    P_{\perp}^{\Theta}.
    \label{eq:pt-effective-mgd}
\end{align}
Since the seed sector is isotropic, the anisotropy of the total configuration
is generated entirely by the additional source,
\begin{equation}
    \Delta
    =
    P_{\perp}-P_r
    =
    P_{\perp}^{\Theta}-P_r^{\Theta}.
    \label{eq:anisotropy-mgd}
\end{equation}

Within the standard MGD prescription, the temporal metric potential is kept
unchanged, while the radial metric component is deformed. Accordingly, the
radial metric potential is written as
\begin{equation}
    e^{-\lambda(r)}
    =
    \mu(r)
    +
    \mathcal{F}(r),
    \label{eq:mgd-deformation}
\end{equation}
where \(\mu(r)\) belongs to the isotropic seed geometry and
\(\mathcal{F}(r)\) denotes the physical geometric deformation induced by the
additional source. The complete interior line element therefore becomes
\begin{equation}
    ds^{2}
    =
    e^{\nu(r)}dt^{2}
    -
    \frac{dr^{2}}{\mu(r)+\mathcal{F}(r)}
    -
    r^{2}d\Omega^{2}.
    \label{eq:mgd-line-element}
\end{equation}
The seed solution is recovered in the limit
\(\mathcal{F}(r)\rightarrow0\).

Substitution of Eq.~\eqref{eq:mgd-deformation} into Einstein's field equations
separates the system into two sets. The first set corresponds to the isotropic
seed sector,
\begin{align}
    8\pi\widehat{\rho}
    &=
    \frac{1-\mu}{r^{2}}
    -
    \frac{\mu'}{r},
    \label{eq:seed-density}
    \\
    8\pi\widehat{P}
    &=
    -\frac{1}{r^{2}}
    +
    \mu
    \left(
        \frac{1}{r^{2}}
        +
        \frac{\nu'}{r}
    \right),
    \label{eq:seed-radial-pressure}
    \\
    8\pi\widehat{P}
    &=
    \frac{\mu}{4}
    \left(
        2\nu''
        +
        \nu'^{2}
        +
        \frac{2\nu'}{r}
    \right)
    +
    \frac{\mu'}{4}
    \left(
        \nu'
        +
        \frac{2}{r}
    \right).
    \label{eq:seed-tangential-pressure}
\end{align}
The equality between the last two equations expresses the isotropic character
of the seed fluid. Once a particular perfect-fluid solution is selected, the
functions $\{\nu,\,\mu,\,\widehat{\rho},\,\widehat{P} \}
    \label{eq:known-seed-functions}$
are completely determined. In the following section, the Tolman IV solution
will be adopted as the isotropic seed.

The second set governs the effective decoupling source,
\begin{align}
    8\pi\rho_{\Theta}
    &=
    -
    \left(
        \frac{\mathcal{F}}{r^{2}}
        +
        \frac{\mathcal{F}'}{r}
    \right),
    \label{eq:theta-density}
    \\
    8\pi P_r^{\Theta}
    &=
    \mathcal{F}
    \left(
        \frac{1}{r^{2}}
        +
        \frac{\nu'}{r}
    \right),
    \label{eq:theta-radial-pressure}
    \\
    8\pi P_{\perp}^{\Theta}
    &=
    \frac{\mathcal{F}}{4}
    \left(
        2\nu''
        +
        \nu'^{2}
        +
        \frac{2\nu'}{r}
    \right)
    +
    \frac{\mathcal{F}'}{4}
    \left(
        \nu'
        +
        \frac{2}{r}
    \right).
    \label{eq:theta-tangential-pressure}
\end{align}
These relations show that the density and both pressure components of the
additional source are completely determined once the deformation
\(\mathcal{F}(r)\) is known.

The total physical variables may therefore be written directly as
\begin{align}
    8\pi\rho
    &=
    8\pi\widehat{\rho}
    -
    \left(
        \frac{\mathcal{F}}{r^{2}}
        +
        \frac{\mathcal{F}'}{r}
    \right),
    \label{eq:total-density-mgd}
    \\
    8\pi P_r
    &=
    8\pi\widehat{P}
    +
    \mathcal{F}
    \left(
        \frac{1}{r^{2}}
        +
        \frac{\nu'}{r}
    \right),
    \label{eq:total-radial-pressure-mgd}
    \\
    8\pi P_{\perp}
    &=
    8\pi\widehat{P}
    +
    \frac{\mathcal{F}}{4}
    \left(
        2\nu''
        +
        \nu'^{2}
        +
        \frac{2\nu'}{r}
    \right)
    +
    \frac{\mathcal{F}'}{4}
    \left(
        \nu'
        +
        \frac{2}{r}
    \right).
    \label{eq:total-tangential-pressure-mgd}
\end{align}
Equations~\eqref{eq:total-density-mgd} and
\eqref{eq:total-tangential-pressure-mgd} make explicit that the complete
anisotropic matter distribution can be obtained directly from the seed
solution and the geometric deformation.

The mass function is also modified by the decoupling sector. Introducing the
seed mass function through
\begin{equation}
    \mu(r)
    =
    1-\frac{2\widehat{m}(r)}{r},
    \label{eq:seed-mass-potential}
\end{equation}
and comparing Eq.~\eqref{eq:mgd-deformation} with
\begin{equation}
    e^{-\lambda(r)}
    =
    1-\frac{2m(r)}{r},
\end{equation}
one obtains
\begin{equation}
    m(r)
    =
    \widehat{m}(r)
    -
    \frac{r}{2}\mathcal{F}(r).
    \label{eq:total-mass-mgd}
\end{equation}
Thus, the deformation modifies both the local mass distribution and the total
gravitational mass of the star.

The contracted Bianchi identities ensure the conservation of the complete
energy--momentum tensor. Under the standard MGD assumption that the two sectors
exchange no energy--momentum except through their common gravitational field,
they may be treated as separately conserved. The seed perfect fluid satisfies
\begin{equation}
    \widehat{P}'
    +
    \frac{\nu'}{2}
    \left(
        \widehat{\rho}+\widehat{P}
    \right)
    =
    0,
    \label{eq:seed-conservation}
\end{equation}
whereas the additional source obeys
\begin{equation}
    \left(P_r^{\Theta}\right)'
    +
    \frac{\nu'}{2}
    \left(
        \rho_{\Theta}+P_r^{\Theta}
    \right)
    -
    \frac{2}{r}
    \left(
        P_{\perp}^{\Theta}-P_r^{\Theta}
    \right)
    =
    0.
    \label{eq:theta-conservation}
\end{equation}
The latter relation is not an independent field equation, since it follows
from Equations~\eqref{eq:theta-density} and \eqref{eq:theta-tangential-pressure}.

\subsection{Closure of the gravitationally decoupled system}

Once the isotropic seed is selected, the functions
\(\nu\), \(\mu\), \(\widehat{\rho}\), and \(\widehat{P}\) are known.
The remaining unknown functions may be expressed as
\begin{equation}
    \left\{
        \mathcal{F},\,
        \rho,\,
        P_r,\,
        P_{\perp}
    \right\},
    \label{eq:unknown-total-functions}
\end{equation}
or, equivalently, in terms of the decoupling sector as
\begin{equation}
    \left\{
        \mathcal{F},\,
        \rho_{\Theta},\,
        P_r^{\Theta},\,
        P_{\perp}^{\Theta}
    \right\}.
    \label{eq:unknown-theta-functions}
\end{equation}
EFEs provide three independent relations, Eqs.~\eqref{eq:theta-density} and \eqref{eq:theta-tangential-pressure}, for these four unknown functions. Therefore, one additional physical condition is required to close the system.

In the present work, closure is achieved by imposing the generalized linear
equation of state introduced in Section~\ref{section-eos} on the total
effective density and radial pressure,
\begin{equation}
    P_r
    =
    \eta\left(\rho-\rho_s\right).
    \label{eq:effective-generalized-eos-mgd}
\end{equation}
Using Eqs.~\eqref{eq:total-density-mgd} and
\eqref{eq:total-radial-pressure-mgd}, this condition becomes
\begin{equation}
    \widehat{P}
    +
    \frac{\mathcal{F}}{8\pi}
    \left(
        \frac{1}{r^{2}}
        +
        \frac{\nu'}{r}
    \right)
    =
    \eta
    \left[
        \widehat{\rho}
        -
        \rho_s
        -
        \frac{1}{8\pi}
        \left(
            \frac{\mathcal{F}}{r^{2}}
            +
            \frac{\mathcal{F}'}{r}
        \right)
    \right].
    \label{eq:eos-deformation-relation}
\end{equation}
It follows that the geometric deformation satisfies the first-order
differential equation
\begin{equation}
    \mathcal{F}'
    +
    \left[
        \frac{\nu'}{\eta}
        +
        \frac{1+\eta}{\eta r}
    \right]\mathcal{F}
    =
    \frac{8\pi r}{\eta}
    \left[
        \eta\left(\widehat{\rho}-\rho_s\right)
        -
        \widehat{P}
    \right].
    \label{eq:deformation-differential-equation}
\end{equation}
All coefficients in this equation are known once the isotropic seed and the
EoS parameters \(\eta\) and \(\rho_s\) have been specified. Regularity at the
center requires
\begin{equation}
    \mathcal{F}(r)=\mathcal{O}(r^{2})
    \qquad
    \text{as}
    \qquad
    r\rightarrow0.
    \label{eq:deformation-central-condition}
\end{equation}
Solving Eq.~\eqref{eq:deformation-differential-equation} determines
\(\mathcal{F}(r)\). The final density, radial pressure, and tangential pressure
then follow directly from
Eqs.~\eqref{eq:total-density-mgd} and 
\eqref{eq:total-tangential-pressure-mgd}.

Therefore, once the Tolman IV geometry \cite{tolman1939static} is adopted as the isotropic seed, the metric potentials and the corresponding seed density and pressure are known up to the constants characterizing the solution. The remaining functional freedom of the gravitationally decoupled configuration is encoded in the geometric deformation $\mathcal{F}(r)$. The three quasi-Einstein equations express the density, radial pressure, and tangential pressure of the additional sector in
terms of this deformation, while the generalized linear equation of state,
imposed on the total effective density and radial pressure, supplies the
independent physical relation required for closure. The resulting first-order
differential equation for $\mathcal{F}(r)$ therefore constitutes the central
structure equation of the present construction.

This closure does not, however, complete the stellar problem by itself. The
deformation equation must be solved subject to central regularity, after which
the total density and pressure profiles are reconstructed from the EFEs. The stellar surface is subsequently determined by the vanishing of the radial pressure, and the constants of the Tolman IV seed are constrained by the matching conditions with the exterior Schwarzschild geometry. The resulting configurations must finally be examined for physical acceptability and dynamical stability. Thus, once the Tolman IV temporal potential has been chosen, the total-fluid radial EoS closes the remaining deformation problem without prescribing an additional anisotropy profile. The choice of temporal potential remains an essential geometric assumption of the construction.

\section{Anisotropic self-bound strange-star configurations}
\label{section-strange-star-model}
\label{section-physical-analysis}

We specialize the construction of Section~\ref{section-mgd} to the Tolman IV temporal potential and impose the generalized linear EoS on the complete matter distribution. The resulting solution can be written in closed form for the four stiffness values $\eta=1,1/2,1/3,1/4$. For each stiffness we retain three compactness values and resolve the dependence of the metric, matter variables, and profile diagnostics on compactness. Constant radial-sound-speed profiles are omitted; the constant $\rho-P_r$ figure is also omitted for $\eta=1$. The surface density is fixed by regularity and Schwarzschild matching once the compactness and length scale are specified. This construction provides the equilibrium backgrounds for the explicitly closed radial-perturbation problem studied in Section~\ref{rpnum:section}.

\subsection{Geometry and dimensionless variables}

Let
\begin{equation}
 x=\frac{r}{R},\quad u=\frac{M}{R},\quad
 a=1-3u,\quad b(x)=a+ux^2.
 \label{eq:s5-variables}
\end{equation}
The temporal potential takes the form
\begin{equation}
 e^{\nu(r)}=b(x)=1-3u+ux^2.
 \label{eq:s5-temporal}
\end{equation}
At the surface, the temporal metric potential satisfies
$e^{\nu(R)}=1-2u$. Keeping it positive and nonsingular throughout the
interior restricts the compactness to $0<u<1/3$. This condition ensures
regularity of the temporal potential, but the matter variables and
stability must still be examined before the configuration can be
considered physically acceptable.

A convenient Tolman IV seed with this temporal potential is \cite{tolman1939static}
\begin{equation}
 \mu(x)=\frac{(1-ux^2)b(x)}{a+2ux^2}.
 \label{eq:s5-seed}
\end{equation}
In the conventional parametrization $e^\nu=B^2(1+r^2/A^2)$, this corresponds to $A^2/R^2=a/u$, $B^2=a$, and $C^2/R^2=1/u$. We denote the complete radial potential by
\begin{equation}
 H(x)=e^{-\lambda(r)}=\mu(x)+\mathcal F(x).
 \label{eq:s5-H}
\end{equation}
The seed choice fixes the decomposition into $\mu$ and $\mathcal F$; the total potential $H$ will be determined directly by the total-fluid EoS. No additional seed-shape parameter is needed for this matched family.

Introduce the dimensionless matter variables
\begin{align}
 \varepsilon&=8\pi R^2\rho,\quad p=8\pi R^2P_r,\nonumber\\
 t&=8\pi R^2P_\perp,\quad s=8\pi R^2\rho_s.
 \label{eq:s5-scaled-matter}
\end{align}
Here $s$ is constant within each star and the EoS becomes $p=\eta(\varepsilon-s)$. With $H_x=dH/dx$, Einstein's equations give
\begin{align}
 \varepsilon&=\frac{1-H}{x^2}-\frac{H_x}{x},
 \label{eq:s5-density}\\
 p&=\frac{H-1}{x^2}+\frac{2uH}{b},
 \label{eq:s5-pr}\\
 t&=\frac{u(2a+ux^2)H}{b^2}
   +\frac{a+2ux^2}{2xb}H_x.
 \label{eq:s5-pt}
\end{align}
The apparent divisions by $x$ are evaluated by their regular limits at the center. The dimensionless anisotropy is $\delta=t-p=8\pi R^2\Delta$.

\subsection{Exact closure and surface matching}

Substituting Eqs.~\eqref{eq:s5-density} and \eqref{eq:s5-pr} into the EoS yields
\begin{equation}
 \eta xH_x+\left(1+\eta+\frac{2ux^2}{b}\right)H
 =1+\eta-\eta sx^2.
 \label{eq:s5-ode}
\end{equation}
This is equivalent to the deformation equation of Section~\ref{section-mgd}, with $\mathcal F=H-\mu$. Set $n=1/\eta$ and define
\begin{equation}
 I_j(x)=\int_0^x y^{n+j}b(y)^n\,dy,\qquad j=0,2.
 \label{eq:s5-integrals}
\end{equation}
The regular solution is
\begin{equation}
 H(x)=\frac{(n+1)I_0(x)-sI_2(x)}{x^{n+1}b(x)^n}.
 \label{eq:s5-regular-H}
\end{equation}
The omitted homogeneous term is proportional to $x^{-(n+1)}b^{-n}$ and is singular at the center. Thus regularity removes its integration constant.

At $x=1$, Schwarzschild matching requires $H(1)=1-2u$. Equation~\eqref{eq:s5-pr} then gives $p(1)=0$, and the EoS gives $\varepsilon(1)=s$. Consequently,
\begin{equation}
 s=\frac{(n+1)I_0(1)-(1-2u)^{n+1}}{I_2(1)}.
 \label{eq:s5-surface-scale}
\end{equation}
The interior density consequently approaches a finite value $\rho(R^-)=\rho_s>0$, whereas the exterior Schwarzschild region has $\rho(R^+)=0$. This finite surface-density jump is the self-bound boundary assumed here. The surface is defined by vanishing radial pressure, not by vanishing density. In particular, for $\eta=1$ one has $\rho=P_r+\rho_s$, so a nonzero endpoint of the density profile at $x=1$ is required by the EoS rather than a failure of the matching conditions.

The quantities $u$, $R$, and $\rho_s$ are therefore not independent. For a chosen positive physical surface density,
\begin{equation}
 R=\sqrt{\frac{s(u,\eta)}{8\pi\rho_s}},\qquad M=uR.
 \label{eq:s5-scale}
\end{equation}
This scaling is essential when interpreting a sequence at fixed EoS: setting $R=1$ is a dimensionless plotting convention, not a common dimensional radius for all stars at fixed $\rho_s$.

For the four cases considered here, $n=1,2,3,4$ is an integer. Define the finite polynomials
\begin{equation}
 L_j(x)=\sum_{k=0}^{n}\binom{n}{k}
 \frac{a^{n-k}u^k x^{2k}}{n+j+2k+1},\qquad j=0,2.
 \label{eq:s5-polynomials}
\end{equation}
Since $I_j=x^{n+j+1}L_j$, the closed solution is simply
\begin{align}
 H(x)&=\frac{(n+1)L_0(x)-sx^2L_2(x)}{b(x)^n},
 \label{eq:s5-polynomial-H}\\
 s&=\frac{(n+1)L_0(1)-(1-2u)^{n+1}}{L_2(1)}.
 \label{eq:s5-polynomial-s}
\end{align}
These expressions specify all four families without long expanded rational functions or numerical integration constants. For example, at $\eta=1$,
\begin{align}
 s&=\frac{6u(3-8u)}{3-7u},\nonumber\\
 H&=\frac{a+(u/2-sa/4)x^2-(su/6)x^4}{b}.
 \label{eq:s5-stiff-example}
\end{align}
The existence of this formal solution does not ensure that it passes the additional matter diagnostics discussed below.

\subsection{Profile definitions and compactness sets}

The configuration sets are listed in Table~\ref{tab:s5-sets}. For each stiffness, nine profiles are grouped into a multipanel figure: $e^{-\lambda}$, $\rho$, $P_r$, $P_t$, $\Delta$, $\rho-P_t$, $\rho-P_r-2P_t$, $v_t$, and $\rho\prime\prime$. The quantity $\rho-P_r$ is not plotted, since its nonnegativity follows directly from Eq.~\eqref{eq:radial-dominant-energy-eos}; for $\eta=1$ it is also constant. Here $P_t\equiv P_\perp$, $\Delta=P_t-P_r$, and primes denote derivatives with respect to the areal radius $r$. All three compactness values in a row of the table are used in every equilibrium-profile panel for that family.

\begin{table}[t]
\centering
\caption{Compactness values used for the equilibrium profiles and radial pulsations. All profile curves are solid; blue, orange, and green denote $u_1$, $u_2$, and $u_3$, respectively.}
\label{tab:s5-sets}
\begin{tabular}{cccc}
\toprule
$\eta$ & $u_1$ & $u_2$ & $u_3$\\
\midrule
$1$ &0.100&0.148&0.196\\
$1/2$ &0.227&0.243&0.259\\
$1/3$ &0.181&0.213&0.244\\
$1/4$ &0.158&0.200&0.241\\
\bottomrule
\end{tabular}
\end{table}

To retain a direct comparison with the dimensionless convention $R=1$, the matter profiles display $R^2\rho$, $R^2P_r$, $R^2P_t$, and the corresponding stress combinations, rather than normalizing each curve by a different surface density. In terms of Eqs.~\eqref{eq:s5-density} and \eqref{eq:s5-pt}, these are $\varepsilon/(8\pi)$, $p/(8\pi)$, and $t/(8\pi)$. The curvature figure displays
\begin{equation}
 R^4\rho''(r)=\frac{1}{8\pi}\frac{d^2\varepsilon}{dx^2}.
 \label{eq:s5-rho-curvature}
\end{equation}
Thus the factors of $R$ restore the dimensions suppressed by setting $R=1$. The different curves do not assume a common dimensional radius at fixed $\rho_s$; such a physical sequence must use Eq.~\eqref{eq:s5-scale}.

\begin{figure}[!htbp]
\centering
\includegraphics[width=\columnwidth]{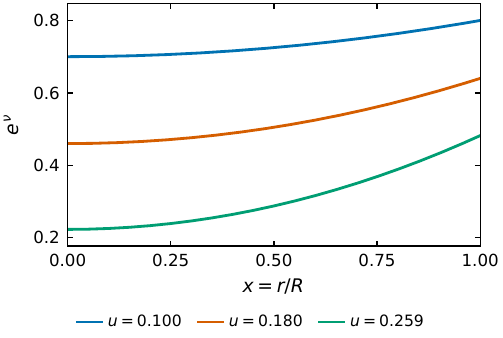}
\caption{Temporal potential $e^\nu$ for $u=0.100,0.180,0.259$. This expression is independent of $\eta$.}
\label{fig:s5-temporal}
\end{figure}

\begin{figure}[!htbp]
\centering
\includegraphics[width=\columnwidth]{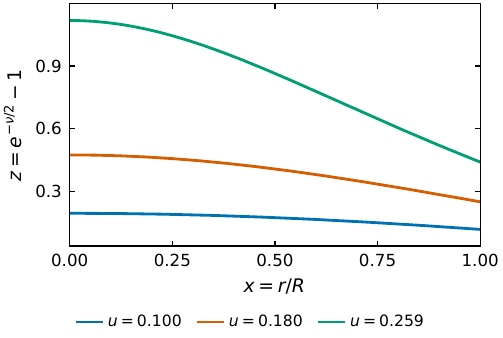}
\caption{Gravitational redshift for the same compactness values as Fig.~\ref{fig:s5-temporal}.}
\label{fig:s5-redshift}
\end{figure}

The common temporal potential and redshift are shown separately in Figs.~\ref{fig:s5-temporal} and \ref{fig:s5-redshift} for $u=0.100,0.180,0.259$. Their expressions are independent of $\eta$ and remain unchanged when the total-fluid EoS is imposed on the radial potential:
\begin{equation}
 e^\nu=b(x),\qquad z(x)=b(x)^{-1/2}-1.
 \label{eq:s5-redshift}
\end{equation}
The temporal potential increases outward, while the redshift decreases. Increasing $u$ lowers $e^\nu$ and raises $z$ throughout the star, with a larger central-to-surface contrast.

We define the radial sound speed and the plotted tangential-profile diagnostic by
\begin{align}
 v_r&=\sqrt{\frac{dP_r}{d\rho}}=\sqrt{\eta},\nonumber\\
 v_t&=\sqrt{\frac{P_t'}{\rho'}}=\sqrt{\frac{t_x}{\varepsilon_x}}.
 \label{eq:s5-velocities}
\end{align}
The radial EoS therefore requires horizontal $v_r$ profiles. At a fixed $\eta$ the three $u$ curves coincide exactly, so these constant profiles are omitted. The constant value differs between stiffnesses. Radially varying $v_r$ curves would contradict Eq.~\eqref{eq:generalized-linear-eos} with constant $\eta$ and $\rho_s$. Here, $v_t$ is calculated from the radial gradients of the equilibrium pressure and density. To determine whether it represents a propagation speed, we also need a constitutive law describing how the anisotropic stress responds to perturbations. At the center, the ratio is evaluated by its regular limit,
\begin{equation}
 v_t^2(0)=\frac{t_{xx}(0)}{\varepsilon_{xx}(0)}
 =\frac{15\eta-1}{10}.
 \label{eq:s5-central-slope}
\end{equation}
The squared tangential ratios are positive for all displayed profiles, so all square roots plotted here are real.

The three stress-combination profiles must also be distinguished. The first two test the upper-pressure inequalities $P_r\leq\rho$ and $P_t\leq\rho$. When tension is present, the dominant energy condition additionally requires $\rho+P_t\geq0$; equivalently one must check $\rho\geq|P_t|$. The third combination,
\begin{equation}
 \mathcal T=\rho-P_r-2P_t,
 \label{eq:s5-trace}
\end{equation}
is a trace combination, not the strong-energy-condition sum. The latter contains $\rho+P_r+2P_t$, together with $\rho+P_i\geq0$. A negative $\mathcal T$ must therefore not be described as a violation of that energy condition. Likewise, $\rho''$ measures the curvature of the density profile; its sign alone does not establish dynamical or convective stability.

\subsection{The maximally stiff family: $\eta=1$}

\begin{figure*}[!t]
\centering

\subfloat[\label{fig:eq-1-H}]{%
  \includegraphics[width=0.32\textwidth]{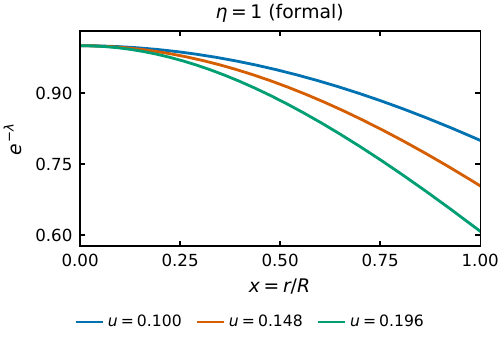}}
\hfill
\subfloat[\label{fig:eq-1-rho}]{%
  \includegraphics[width=0.32\textwidth]{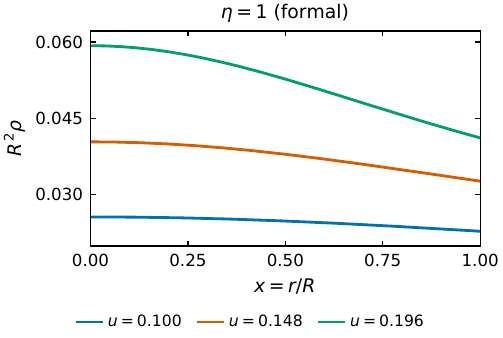}}
\hfill
\subfloat[\label{fig:eq-1-pr}]{%
  \includegraphics[width=0.32\textwidth]{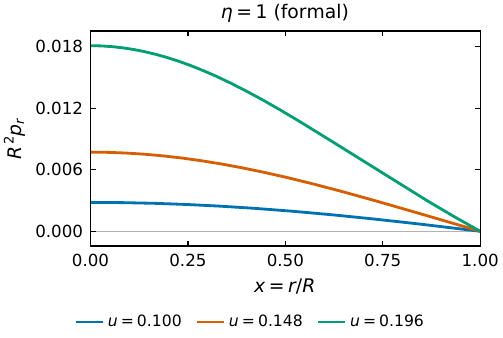}}

\par\medskip

\subfloat[\label{fig:eq-1-pt}]{%
  \includegraphics[width=0.32\textwidth]{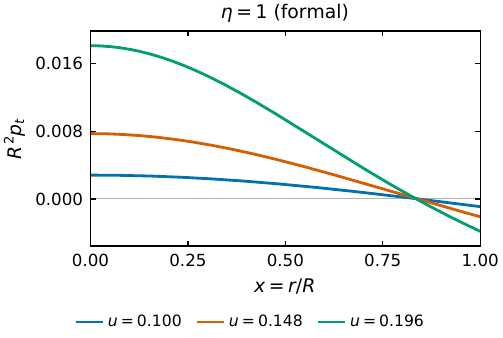}}
\hfill
\subfloat[\label{fig:eq-1-delta}]{%
  \includegraphics[width=0.32\textwidth]{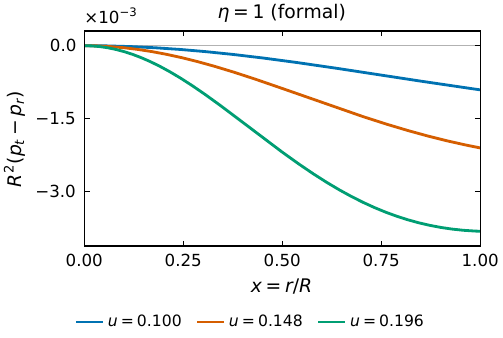}}
\hfill
\subfloat[\label{fig:eq-1-rho_pt}]{%
  \includegraphics[width=0.32\textwidth]{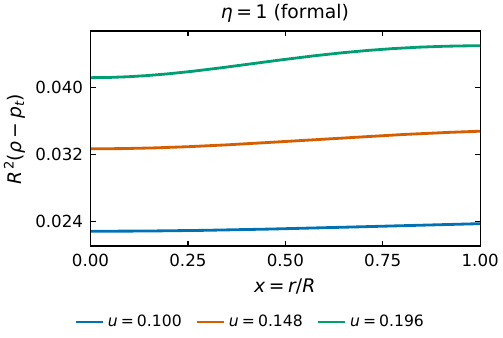}}

\par\medskip

\subfloat[\label{fig:eq-1-trace}]{%
  \includegraphics[width=0.32\textwidth]{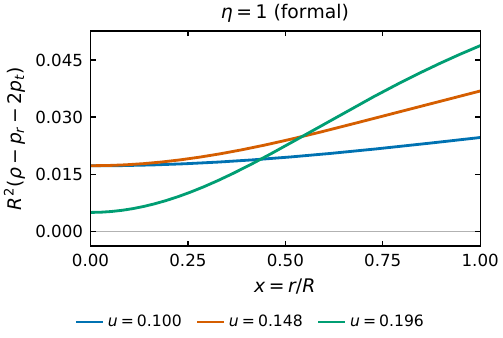}}
\hfill
\subfloat[\label{fig:eq-1-vt}]{%
  \includegraphics[width=0.32\textwidth]{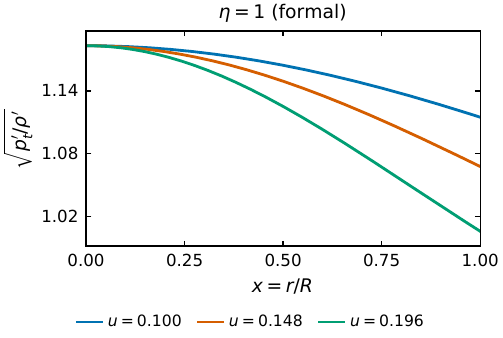}}
\hfill
\subfloat[\label{fig:eq-1-rho2}]{%
  \includegraphics[width=0.32\textwidth]{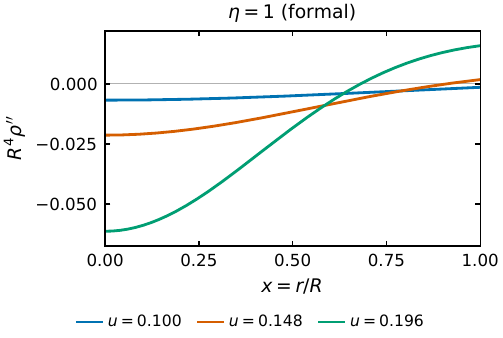}}

\caption{Equilibrium profiles and diagnostics for $\eta=1$:
(a) radial metric potential $e^{-\lambda}$;
(b) density $R^2\rho$, with finite positive surface density;
(c) radial pressure $R^2P_r$, which vanishes at $r=R$;
(d) tangential pressure $R^2P_t$;
(e) anisotropy $R^2(P_t-P_r)$;
(f) tangential stress combination $R^2(\rho-P_t)$;
(g) trace combination $R^2(\rho-P_r-2P_t)$, which is not the
strong-energy-condition sum;
(h) tangential-profile diagnostic
$v_t=\sqrt{P_t^{\prime}/\rho^{\prime}}$;
and (i) density curvature $R^4\rho^{\prime\prime}$,
whose sign changes indicate inflection points.}
\label{fig:eta1-profiles}

\end{figure*}

Figures~\ref{fig:eq-1-H}--\ref{fig:eq-1-rho2} show the nine retained profiles for $u=0.100,0.148,0.196$. The radial metric is regular at the center, $H(0)=1$, and reaches $H(1)=1-2u$. The density and radial pressure decrease outward, with $P_r(R)=0$ and a finite positive surface density. The central density and pressure grow across the displayed compactness set. The tangential pressure equals the radial pressure at the center and becomes negative in an outer layer: its zero occurs at $x\simeq0.8413,0.8388,0.8359$, respectively. The anisotropy is negative for $0<x\leq1$, corresponding to an inward anisotropic contribution to hydrostatic balance.

The relation $\rho-P_r=\rho_s$ at $\eta=1$ is an exact EoS consequence, so its constant figure is omitted. The combinations $\rho-P_t$ and $\mathcal T$ remain positive for the displayed configurations. The radial diagnostic is $v_r=1$, whereas $v_t(0)=\sqrt{7/5}\simeq1.1832$. In fact, $v_t>1$ throughout all three displayed interiors. These solutions are therefore retained as formal comparisons and fail the usual subluminal tangential-profile diagnostic; they are not presented as configurations satisfying every physical-acceptability criterion.

The density curvature is negative throughout the $u=0.100$ profile. For $u=0.148$ and $0.196$ it changes sign at $x\simeq0.9104$ and $0.6810$. These inflection points do not reverse the outward decrease of $\rho$.

\subsection{The intermediate-stiffness family: $\eta=1/2$}

\begin{figure*}[!t]
\centering

\subfloat[\label{fig:eq-2-H}]{%
  \includegraphics[width=0.32\textwidth]{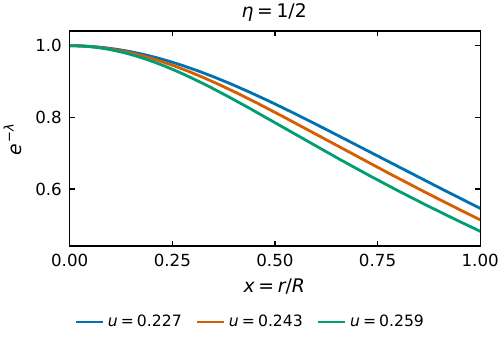}}
\hfill
\subfloat[\label{fig:eq-2-rho}]{%
  \includegraphics[width=0.32\textwidth]{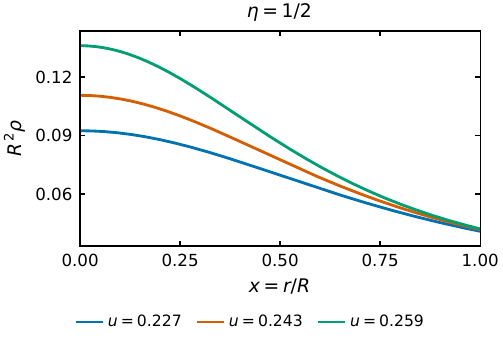}}
\hfill
\subfloat[\label{fig:eq-2-pr}]{%
  \includegraphics[width=0.32\textwidth]{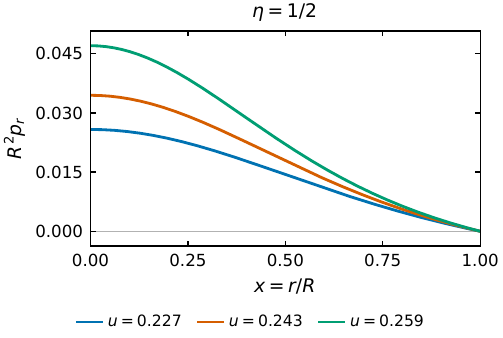}}

\par\medskip

\subfloat[\label{fig:eq-2-pt}]{%
  \includegraphics[width=0.32\textwidth]{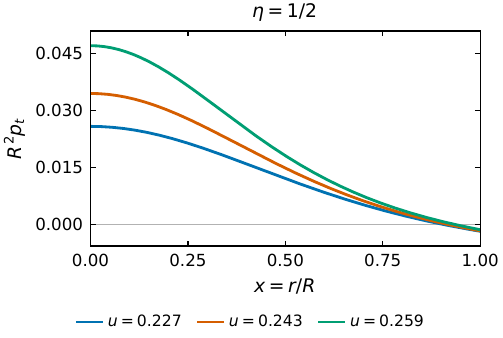}}
\hfill
\subfloat[\label{fig:eq-2-delta}]{%
  \includegraphics[width=0.32\textwidth]{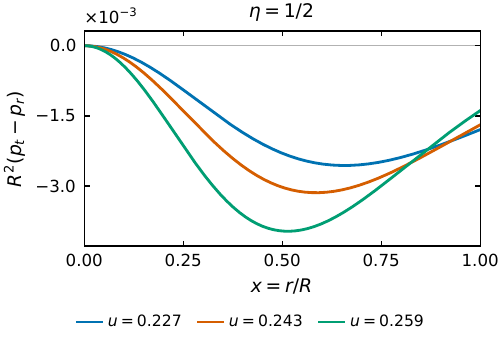}}
\hfill
\subfloat[\label{fig:eq-2-rho_pt}]{%
  \includegraphics[width=0.32\textwidth]{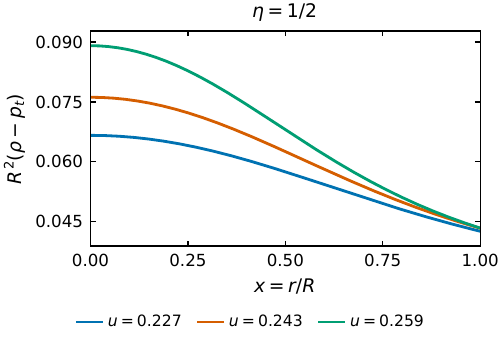}}

\par\medskip

\subfloat[\label{fig:eq-2-trace}]{%
  \includegraphics[width=0.32\textwidth]{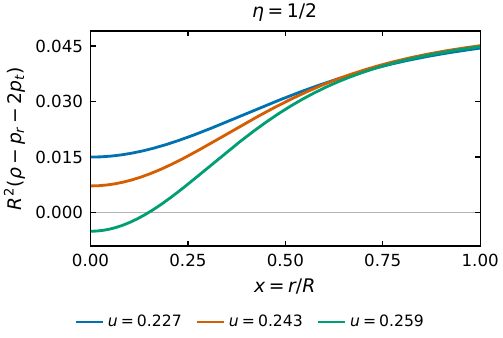}}
\hfill
\subfloat[\label{fig:eq-2-vt}]{%
  \includegraphics[width=0.32\textwidth]{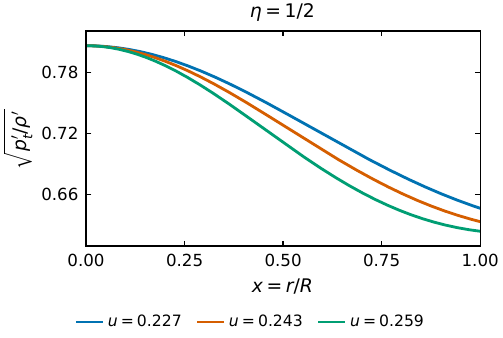}}
\hfill
\subfloat[\label{fig:eq-2-rho2}]{%
  \includegraphics[width=0.32\textwidth]{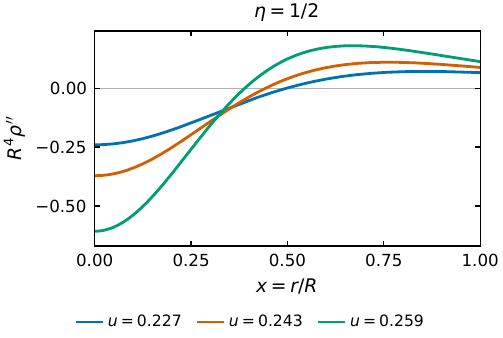}}

\caption{Equilibrium profiles and diagnostics for $\eta=1/2$:
(a) radial metric potential $e^{-\lambda}$;
(b) density $R^2\rho$, with finite positive surface density;
(c) radial pressure $R^2P_r$, which vanishes at $r=R$;
(d) tangential pressure $R^2P_t$;
(e) anisotropy $R^2(P_t-P_r)$;
(f) tangential stress combination $R^2(\rho-P_t)$;
(g) trace combination $R^2(\rho-P_r-2P_t)$, which is not the
strong-energy-condition sum;
(h) tangential-profile diagnostic
$v_t=\sqrt{P_t^{\prime}/\rho^{\prime}}$;
and (i) density curvature $R^4\rho^{\prime\prime}$,
whose sign changes indicate inflection points.}
\label{fig:eta2-profiles}
\end{figure*}

Figures~\ref{fig:eq-2-H}--\ref{fig:eq-2-rho2} retain $u=0.227,0.243,0.259$. The radial metric remains positive, and the density and both pressures decrease outward. The radial pressure vanishes at the surface, while the tangential pressure crosses zero at $x\simeq0.9051,0.9169,0.9350$ and is negative beyond those radii. The anisotropy is negative away from the center for all three curves. Hence these configurations contain a surface-adjacent region of tangential tension; imposing $P_t\geq0$ as an additional condition would exclude them.

Both $\rho-P_r$ and $\rho-P_t$ remain positive. The trace combination is positive for the first two compactness values, but the $u=0.259$ configuration has $\mathcal T<0$ in its central region, up to $x\simeq0.1485$. This is reported explicitly rather than suppressing the curve. The independent energy-condition checks below show why it must not be confused with a violation of the strong energy condition.

The radial velocity is the common constant $v_r=1/\sqrt2\simeq0.7071$. The tangential diagnostic starts at $\sqrt{13/20}\simeq0.8062$ and decreases to approximately $0.6462$, $0.6331$, and $0.6236$ at the respective surfaces. The curvature $\rho''$ changes sign at $x\simeq0.4971,0.4428,0.3891$: the inflection point moves inward as compactness increases, although the density remains monotone.

\subsection{The MIT bag radial relation: $\eta=1/3$}

\begin{figure*}[!t]
\centering

\subfloat[\label{fig:eq-3-H}]{%
  \includegraphics[width=0.32\textwidth]{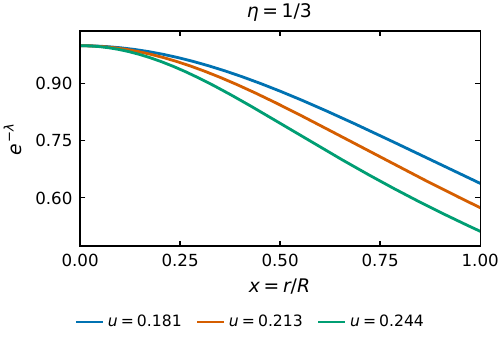}}
\hfill
\subfloat[\label{fig:eq-3-rho}]{%
  \includegraphics[width=0.32\textwidth]{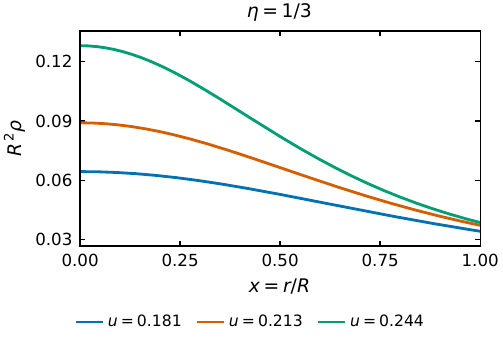}}
\hfill
\subfloat[\label{fig:eq-3-pr}]{%
  \includegraphics[width=0.32\textwidth]{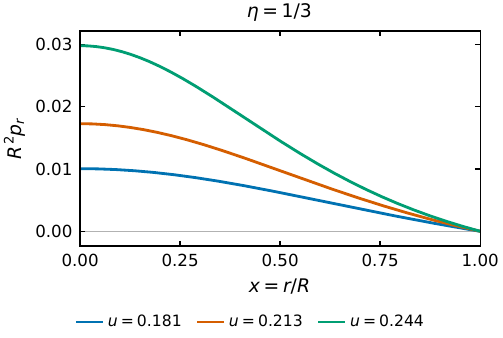}}

\par\medskip

\subfloat[\label{fig:eq-3-pt}]{%
  \includegraphics[width=0.32\textwidth]{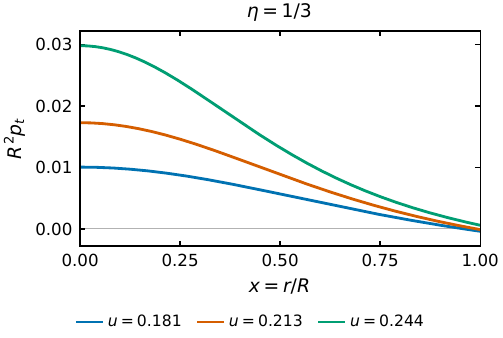}}
\hfill
\subfloat[\label{fig:eq-3-delta}]{%
  \includegraphics[width=0.32\textwidth]{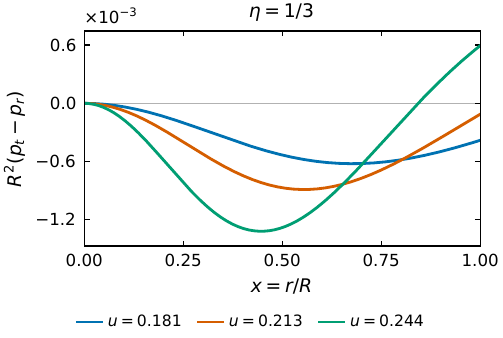}}
\hfill
\subfloat[\label{fig:eq-3-rho_pt}]{%
  \includegraphics[width=0.32\textwidth]{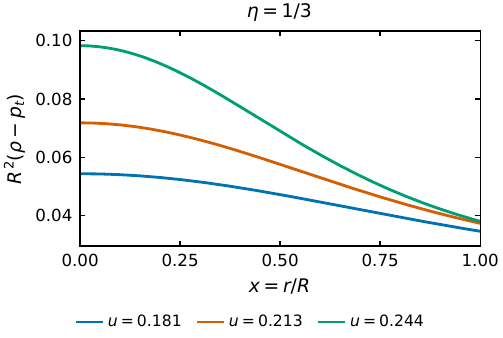}}

\par\medskip

\subfloat[\label{fig:eq-3-trace}]{%
  \includegraphics[width=0.32\textwidth]{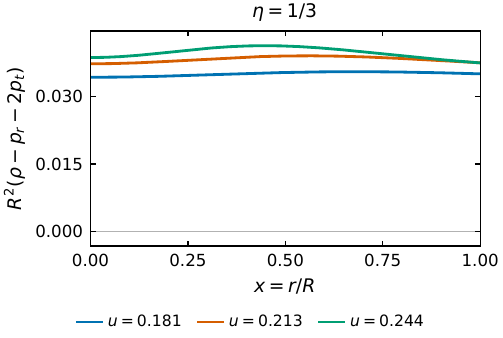}}
\hfill
\subfloat[\label{fig:eq-3-vt}]{%
  \includegraphics[width=0.32\textwidth]{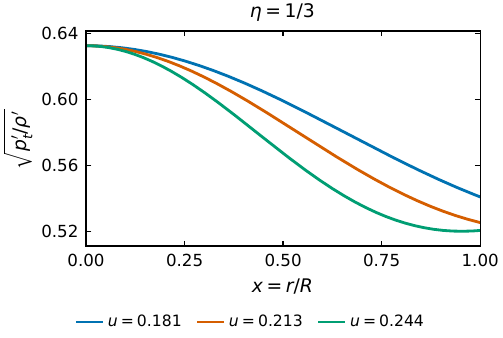}}
\hfill
\subfloat[\label{fig:eq-3-rho2}]{%
  \includegraphics[width=0.32\textwidth]{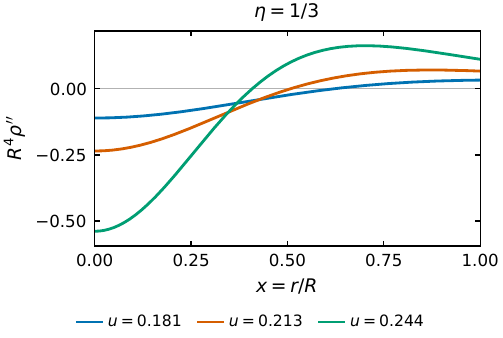}}

\caption{Equilibrium profiles and diagnostics for $\eta=1/3$:
(a) radial metric potential $e^{-\lambda}$;
(b) density $R^2\rho$, with finite positive surface density;
(c) radial pressure $R^2P_r$, which vanishes at $r=R$;
(d) tangential pressure $R^2P_t$;
(e) anisotropy $R^2(P_t-P_r)$;
(f) tangential stress combination $R^2(\rho-P_t)$;
(g) trace combination $R^2(\rho-P_r-2P_t)$, which is not the
strong-energy-condition sum;
(h) tangential-profile diagnostic
$v_t=\sqrt{P_t^{\prime}/\rho^{\prime}}$;
and (i) density curvature $R^4\rho^{\prime\prime}$,
whose sign changes indicate inflection points.}
\label{fig:eta3-profiles}
\end{figure*}

For $\eta=1/3$, identifying $\rho_s=4B$ gives $P_r=(\rho-4B)/3$. Figures~\ref{fig:eq-3-H}--\ref{fig:eq-3-rho2} use $u=0.181,0.213,0.244$. Again $H>0$, the density and radial pressure decrease outward, and the radial pressure reaches zero at a nonzero surface density. The first two configurations have small tangential tensions near the surface, beginning at $x\simeq0.9597$ and $0.9904$. At $u=0.244$, $P_t$ remains positive throughout; the anisotropy changes from negative to positive at $x\simeq0.8433$. Thus the sign of the anisotropy and the sign of the tangential pressure carry different information.

All three displayed stress combinations are positive. The radial velocity is $v_r=1/\sqrt3\simeq0.5774$, and the tangential diagnostic starts at $\sqrt{2/5}\simeq0.6325$ and remains below unity. The density-curvature zeros occur at $x\simeq0.6191,0.5073,0.4084$. If nonnegative tangential pressure and the subluminal profile diagnostics are required simultaneously, only the $u=0.244$ example among these three passes those particular tests. This statement concerns the sampled configurations and is not a bound on the complete family.

\subsection{The softer family: $\eta=1/4$}

\begin{figure*}[!t]
\centering

\subfloat[\label{fig:eq-4-H}]{%
  \includegraphics[width=0.32\textwidth]{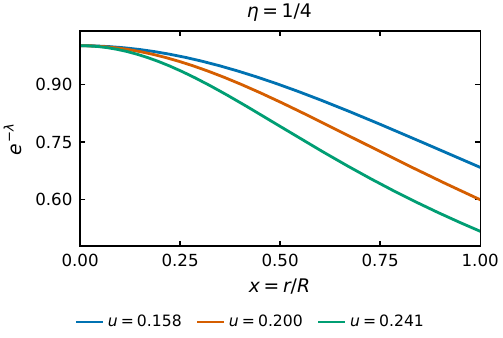}}
\hfill
\subfloat[\label{fig:eq-4-rho}]{%
  \includegraphics[width=0.32\textwidth]{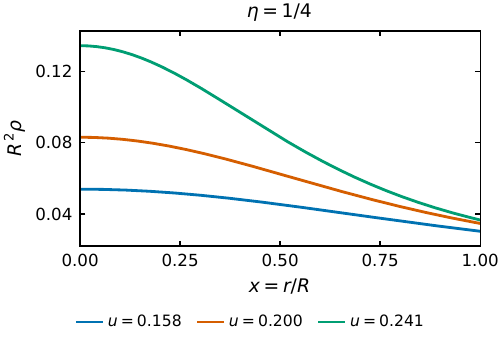}}
\hfill
\subfloat[\label{fig:eq-4-pr}]{%
  \includegraphics[width=0.32\textwidth]{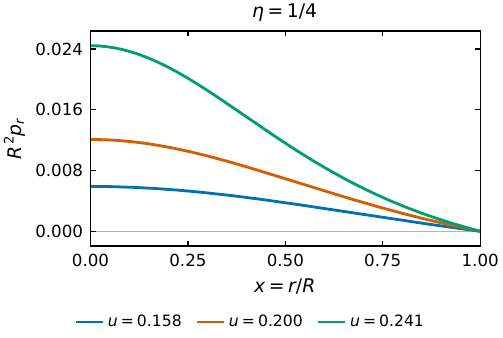}}

\par\medskip

\subfloat[\label{fig:eq-4-pt}]{%
  \includegraphics[width=0.32\textwidth]{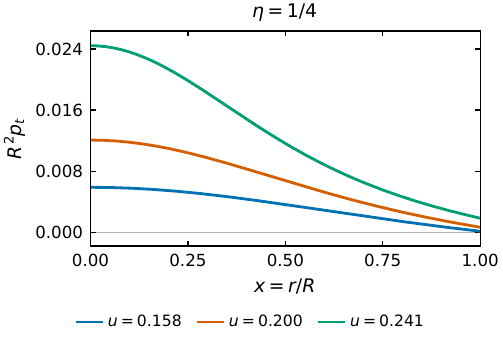}}
\hfill
\subfloat[\label{fig:eq-4-delta}]{%
  \includegraphics[width=0.32\textwidth]{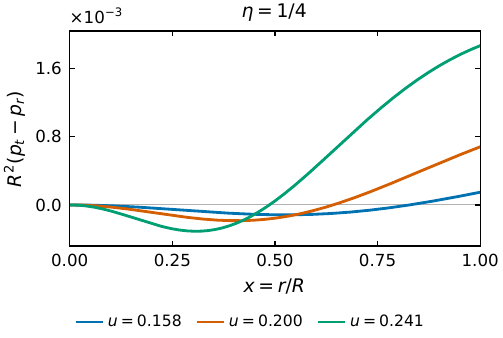}}
\hfill
\subfloat[\label{fig:eq-4-rho_pt}]{%
  \includegraphics[width=0.32\textwidth]{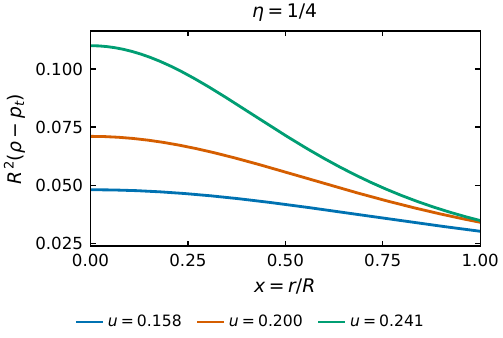}}

\par\medskip

\subfloat[\label{fig:eq-4-trace}]{%
  \includegraphics[width=0.32\textwidth]{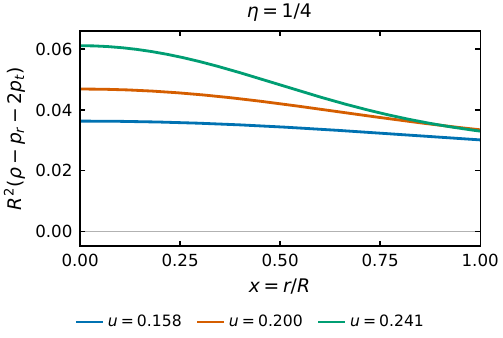}}
\hfill
\subfloat[\label{fig:eq-4-vt}]{%
  \includegraphics[width=0.32\textwidth]{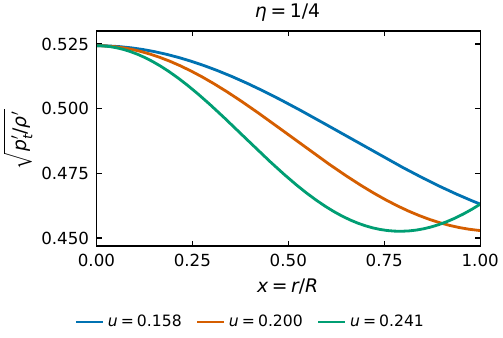}}
\hfill
\subfloat[\label{fig:eq-4-rho2}]{%
  \includegraphics[width=0.32\textwidth]{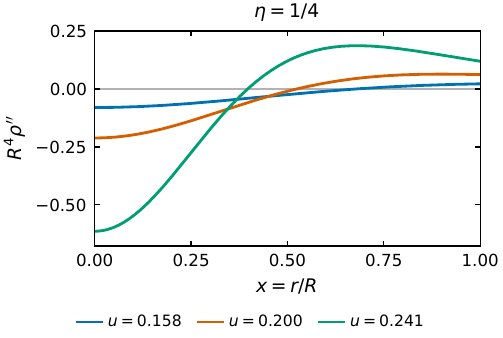}}

\caption{Equilibrium profiles and diagnostics for $\eta=1/4$:
(a) radial metric potential $e^{-\lambda}$;
(b) density $R^2\rho$, with finite positive surface density;
(c) radial pressure $R^2P_r$, which vanishes at $r=R$;
(d) tangential pressure $R^2P_t$;
(e) anisotropy $R^2(P_t-P_r)$;
(f) tangential stress combination $R^2(\rho-P_t)$;
(g) trace combination $R^2(\rho-P_r-2P_t)$, which is not the
strong-energy-condition sum;
(h) tangential-profile diagnostic
$v_t=\sqrt{P_t^{\prime}/\rho^{\prime}}$;
and (i) density curvature $R^4\rho^{\prime\prime}$,
whose sign changes indicate inflection points.}
\label{fig:eta4-profiles}
\end{figure*}

Figures~\ref{fig:eq-4-H}--\ref{fig:eq-4-rho2} use $u=0.158,0.200,0.241$. The radial metric and density are positive, both pressures decrease outward, and $P_t$ stays positive, including at the surface. The anisotropy is initially negative but becomes positive beyond $x\simeq0.8279,0.6417,0.4865$, respectively. A larger compactness therefore moves the change in the anisotropic contribution to smaller fractional radius within this set.

All three configurations have positive values of $\rho-P_r$, $\rho-P_t$,
and $\mathcal T$, and share the constant radial sound speed $v_r=1/2$.
The tangential diagnostic starts at
$\sqrt{11/40}\simeq0.5244$ at the center, independently of $u$, and ranges
from approximately $0.4526$ to $0.5244$ across the three models. Near the surface, the shape of the $v_t$ profile varies with compactness. In the most compact model ($u=0.241$), $v_t$ decreases to a minimum before increasing again towards the surface. The density continues to decrease throughout the interior, with
$\rho'<0$ for $0<x\leq1$, even though its curvature changes sign at
$x\simeq0.6750,0.5232,0.3966$, respectively. The pressures remain
nonnegative and the velocity profiles stay below unity in all three
cases. These checks support the acceptability of the equilibrium
profiles, but dynamical stability requires a separate perturbation
analysis.

\subsection{Consistency checks and physical scope}

All twelve configurations satisfy central isotropy, $P_r(0)=P_t(0)$, and the matching conditions $H(1)=1-2u$, $P_r(R)=0$, and $\rho(R)=\rho_s$. The conservation equation is the exact identity
\begin{equation}
 p_x=-\frac{ux}{b}(\varepsilon+p)+\frac{2(t-p)}{x}.
 \label{eq:s5-balance}
\end{equation}
The EoS, matching, central limits, and stress-balance relations were checked algebraically. Independent integration of Eq.~\eqref{eq:s5-ode} agrees with the closed solution to better than $10^{-8}$ in $H$ for all twelve examples, and integrating the density reproduces $M=uR$.

Evaluation on 4001 equally spaced points gives $H>0$, $\rho>0$, $P_r\geq0$ (up to numerical roundoff at the surface), and decreasing density and radial pressure in every displayed configuration. The inequalities $\rho\geq|P_r|$, $\rho\geq|P_t|$, $\rho+P_i\geq0$, and $\rho+P_r+2P_t\geq0$ also hold on that grid. These numerical checks apply to the displayed examples and do not establish a global allowed interval of $u$. In particular, they neither remove the reported tangential tensions nor make the $\eta=1$ family pass its failed tangential-profile test.

Matching requires $P_r(R)=0$, but not $P_t(R)=0$. Tangential tensions occur in phenomenological anisotropic strange-star models~\cite{das2023anisotropic}. A model that forbids tension must exclude the affected configurations. Self-boundness does not by itself provide a microscopic justification for anisotropic stress.

It should be noted that negative pressure anisotropy, $\Delta < 0$, has previously been considered in relativistic models of neutron stars and quark stars, including studies on non-radial oscillations, radial stability, and $f$-modes \cite{doneva2012nonradial,pretel2024effects,mondal2025quasinormal}. Pressure anisotropy may arise from condensates, elastic or crystalline phases, and strong magnetic fields; however, its sign depends on the material response and, where relevant, the field geometry. For $\Delta=P_t-P_r<0$, radial pressure dominates over tangential pressure, and the anisotropic contribution to hydrostatic balance, $2\Delta/r$, points inward. The presence of this phenomenology does not, in itself, imply dynamical instability within the star.

The equilibrium profiles alone do not establish dynamical stability. Section~\ref{rpnum:section} specifies two constitutive prescriptions for the anisotropy perturbation and solves the associated radial eigenvalue problems. Those conditional stability results do not supply a microscopic origin for the anisotropic stresses or determine a maximum stable mass.

\FloatBarrier
\section{Radial pulsations of the self-bound configurations}
\label{rpnum:section}
\label{section-radial-pulsations}

We now investigate the linear radial response of the twelve equilibrium
configurations considered above. A static anisotropy profile does not specify
how the anisotropic stress responds to a displacement of the fluid. We therefore
state the material response explicitly and compare two closures that reproduce
the same equilibrium. The resulting stability statements refer to these
closures. This distinction is particularly relevant when the tangential
pressure remains nonzero at the surface, as in radial-stability studies of
anisotropic strange stars~\cite{arbanil2016anisotropic}.

\subsection{Material response and perturbation equations}
\label{rpnum:closure}

Let $\xi(r)e^{i\omega t}$ denote the radial displacement and let $\Delta_L$
and $\delta$ denote Lagrangian and Eulerian perturbations, respectively, so
that $\Delta_L F=\delta F+\xi\,dF/dr$. The equilibrium parameters $\eta$ and
$\rho_s$ are held fixed during each oscillation. Consequently,
\begin{equation}
 \Delta_L P_r=\eta\,\Delta_L\rho.
 \label{rpnum:perturbed-eos}
\end{equation}
Writing $\Sigma=P_t-P_r$ for the physical anisotropy, we introduce a material
label $a_L$ that is transported with each fluid layer and prescribe
\begin{equation}
 \Sigma(a_L,\rho)=\Sigma_0(a_L)
 \left[\frac{\rho+P_r}{\rho_0(a_L)+P_{r0}(a_L)}\right]^\kappa,
 \qquad \kappa=0,1.
 \label{rpnum:material-law}
\end{equation}
The subscript $0$ denotes the chosen background. For $\kappa=0$, each layer
preserves its anisotropy, whereas for $\kappa=1$ it preserves
$\Sigma/(\rho+P_r)$. These are phenomenological response laws; they are not
deduced from the equilibrium equation of state or from gravitational
decoupling. Both preserve the central behavior $\Sigma=O(r^2)$ on the regular
perturbative branch. Their linearization gives
\begin{equation}
 \Delta_L\Sigma=
 \frac{\kappa(1+\eta)\Sigma}{\eta(\rho+P_r)}\Delta_LP_r.
 \label{rpnum:anisotropy-response}
\end{equation}
In particular, $\Delta_L\Sigma=0$ and $\delta\Sigma=0$ are different
prescriptions.

The dynamic response of anisotropic stresses cannot be uniquely determined from the radial equilibrium equation of state. To address this, an additional constitutive prescription is required to close the system of radial perturbations. In this regard, previous studies have demonstrated the importance of anisotropic equations of state and stress prescriptions for determining the radial stability and oscillation spectra of compact stars \cite{dev2003anisotropic,horvat2011radial,arbanil2016anisotropic,panotopoulos2026modeling}. In this work, the choice $\kappa = 0$ represents an anisotropy that is “frozen” in the Lagrangian sense, while $\kappa = 1$ defines the ratio between the anisotropy and the enthalpy as fluid elements undergo displacement. These two prescriptions are adopted as phenomenological limiting cases rather than as derived microscopic laws. Consequently, the resulting frequencies and conclusions regarding stability must be understood as being contingent upon the assumed anisotropic constitutive behaviour. Furthermore, it should be noted that future work could examine the full range $0 \leq \kappa \leq 1$ to determine how the spectrum $\omega_{n}^{2}(\kappa)$ evolves continuously.

For the numerical problem we use $x=r/R$, $H=e^{-\lambda}$,
$b=e^\nu=1-3u+ux^2$, and the dimensionless background variables
\begin{equation}
 \begin{aligned}
 \varepsilon&=8\pi R^2\rho,& p&=8\pi R^2P_r,\\
 \sigma&=8\pi R^2(P_t-P_r),& w&=\varepsilon+p.
 \end{aligned}
 \label{rpnum:background-variables}
\end{equation}
The factors in Eq.~\eqref{rpnum:background-variables} use geometrized units,
$G=c=1$. We define
\begin{equation}
 y=\frac{\xi}{r},\qquad d=8\pi R^2\Delta_LP_r,
 \qquad q=\left(\frac{\omega R}{c_{\rm light}}\right)^2.
 \label{rpnum:variables}
\end{equation}
Primes in this section denote differentiation with respect to $x$.
Linearizing the Einstein constraints and radial stress-energy conservation,
including the metric perturbations, gives
\begin{align}
 y'&=a_y y+B_d d,\label{rpnum:displacement-equation}\\
 d'&=(C_y+qW_y)y+E_d d,\label{rpnum:pressure-equation}
\end{align}
where
\begin{align}
 a_y&=\frac{\nu'}2-\frac3x-\frac{2\sigma}{wx},\nonumber\\
 B_d&=-\frac{1}{\eta wx},\label{rpnum:coefficients-ab}\\
 C_y&=-4p'-\frac{wpx}{H}+\frac{xw\nu'^2}{4}
       +\frac{2\sigma}{x}\nonumber\\
    &\quad+2\sigma\nu'-\frac{4\sigma^2}{wx},\nonumber\\
 W_y&=\frac{wx}{Hb},\label{rpnum:coefficients-cw}\\
 E_d&=-\frac{\nu'}2-\frac{wx}{2H}
       +\frac{2[\kappa(1+\eta)-1]\sigma}{\eta wx}.
 \label{rpnum:coefficient-e}
\end{align}
For example, the metric constraints used in this reduction are
\begin{align}
 \delta\lambda&=-\frac{x^2wy}{H},\nonumber\\
 \delta\nu'&=\left(\nu'+\frac1x\right)\delta\lambda
                +\frac{x\,\delta p}{H},
 \label{rpnum:metric-constraints}
\end{align}
with $\delta p=d-xyp'$ and
$\delta\varepsilon=d/\eta-xy\varepsilon'$. The anisotropy perturbation is
\begin{equation}
 \delta\sigma=\frac{\kappa(1+\eta)\sigma}{\eta w}d-xy\sigma'.
 \label{rpnum:eulerian-anisotropy}
\end{equation}
Thus the calculation does not freeze the metric or infer a dynamical
anisotropy law from the ratio $P_t'/\rho'$ of the equilibrium profiles.

\subsection{Boundary conditions and spectral problem}
\label{rpnum:boundary-conditions}

Regularity at the center fixes the relation between the two perturbation
variables. Choosing the arbitrary linear normalization $y(0)=1$ gives
\begin{equation}
 d(0)=-3\eta w(0),\qquad \xi(0)=0.
 \label{rpnum:center}
\end{equation}
The regular even-power expansion through $x^6$ supplies initial data close
to the center; it is used only to start the integration. The surface moves
with the fluid, and the appropriate free-boundary condition is
\begin{equation}
 \Delta_LP_r(R)=0,\qquad\hbox{or equivalently}\qquad d(1)=0.
 \label{rpnum:surface}
\end{equation}
Neither $\delta P_r(R)=0$ at fixed radius nor $P_t(R)=0$ is imposed.
For the response law above, Eq.~\eqref{rpnum:surface} also implies
$\Delta_L\rho(R)=\Delta_L\Sigma(R)=0$. The nonzero equilibrium surface
density therefore causes no singular coefficient: although
$\Gamma_r=\eta(\rho+P_r)/P_r$ diverges as $P_r\to0$, the combination
$\Gamma_rP_r=\eta(\rho+P_r)$ entering the equations remains finite.

Eliminating $d$ yields
\begin{equation}
 y''=A_2y'+(V_0-qV_1)y,
 \label{rpnum:second-order}
\end{equation}
where
\begin{align}
 A_2&=a_y+E_d+B_d'/B_d,\nonumber\\
 V_0&=a_y'+B_dC_y-a_y(E_d+B_d'/B_d),\nonumber\\
 V_1&=-B_dW_y=\frac{1}{\eta Hb}>0.
 \label{rpnum:sl-coefficients}
\end{align}
With $\mathcal P=\exp[-\int A_2\,dx]>0$,
$\mathcal Q=\mathcal P V_0$ and $\mathcal W=\mathcal P V_1$, this becomes
\begin{equation}
 - (\mathcal P y')'+\mathcal Qy=q\mathcal Wy.
 \label{rpnum:sturm-liouville}
\end{equation}
Here $\mathcal P\sim x^4$ at the center; the irregular solution
$y\sim x^{-3}$ is excluded. At the surface,
$y'(1)=a_y(1)y(1)$. Accordingly, the Rayleigh quotient includes the
surface term,
\begin{equation}
 q=\frac{\displaystyle\int_0^1
       (\mathcal P y'^2+\mathcal Qy^2)\,dx
       -\mathcal P(1)a_y(1)y(1)^2}
       {\displaystyle\int_0^1\mathcal Wy^2\,dx}.
 \label{rpnum:rayleigh}
\end{equation}
For this real conservative problem, the fundamental mode is nodeless.
The condition $q_0>0$ implies linear radial stability under the specified
closure, whereas $q_0<0$ corresponds to exponential growth.

\subsection{Numerical spectra and mode profiles}
\label{rpnum:results}
\label{section-results}

We computed the fundamental mode and the first three radial overtones for
each of the twelve backgrounds and both values of $\kappa$, obtaining
96 eigenvalues. A continuous Pr\"ufer-phase search includes the negative
$q$ region and is followed by shooting refinement of
Eqs.~\eqref{rpnum:displacement-equation} and \eqref{rpnum:pressure-equation}.
For every background and closure, positivity checks with exact rational
Bernstein coefficients establish $V_0/V_1+100\geq0$ on $0\leq x\leq1$,
and $a_y(1)\leq0$. Equation~\eqref{rpnum:rayleigh} then gives the lower
bound $q\geq-100$, which sets the lower end of the spectral search.

The displacement eigenfunctions have respectively zero, one, two and three
interior nodes. Changing arithmetic precision and the central starting
point changes the eigenvalues by less than $2.4\times10^{-17}$ in the
recorded comparisons. The maximum normalized surface residual is
$6.0\times10^{-23}$; agreement with the Rayleigh quotient and the maximum
normalized overlap of distinct modes are better than
$2.8\times10^{-18}$ and $5.1\times10^{-18}$, respectively. An independent
implementation of the linearized Einstein and conservation equations
agrees within $5.7\times10^{-14}$. The latter refines the previously
located roots and is not an independent global spectral search. These
figures quantify numerical consistency, not physical uncertainty or
rigorous interval bounds on the eigenvalues.

Table~\ref{rpnum:fundamentals} and the fundamental-mode figures below
give the fundamental eigenvalues. All 24 background--closure combinations
have $q_0>0$. Within each sampled family and each closure, $q_0$ decreases
as $u$ increases. The smallest value is $q_0=0.09244687$ for
$\eta=1/4$, $u=0.241$ and $\kappa=0$; its sign remains positive.
The effect of changing the material response is not universal:
$\kappa=1$ lowers $q_0$ in the sampled $\eta=1,1/2,1/3$ families,
whereas for $\eta=1/4$ it lowers $q_0$ at $u=0.158$ and raises it at
$u=0.200$ and $0.241$. No stability boundary is extrapolated from
these three-point samples. For example, at $\eta=1/4$ and $u=0.241$, changing from $\kappa=0$ to $\kappa=1$ increases $q_0$ from $0.09244687$ to $0.11232334$, a relative increase of approximately $21.5\%$. Because the background and its radius are unchanged, the corresponding fundamental frequency increases by approximately $10.2\%$, as follows from $f\propto\sqrt{q_0}$. This quantifies the sensitivity of the radial spectrum to the adopted anisotropic response even when the equilibrium configuration is fixed.

\begin{table}[t]
 \centering
 \caption{Fundamental eigenvalue $q_0=(\omega_0R/c)^2$ for the twelve equilibrium configurations and the two material responses in Eq.~\eqref{rpnum:material-law}.}
 \label{rpnum:fundamentals}
 \begin{tabular}{cccc}
 \hline
 $\eta$ & $u$ & $q_0\ (\kappa=0)$ & $q_0\ (\kappa=1)$\\
 \hline
 $1$   & 0.100 & 5.89513198 & 5.72364703\\
 $1$   & 0.148 & 4.28679684 & 4.05634429\\
 $1$   & 0.196 & 2.87569435 & 2.60583884\\
 $1/2$ & 0.227 & 0.82625953 & 0.70995096\\
 $1/2$ & 0.243 & 0.62339139 & 0.51179733\\
 $1/2$ & 0.259 & 0.43076829 & 0.32740594\\
 $1/3$ & 0.181 & 0.85825179 & 0.81078068\\
 $1/3$ & 0.213 & 0.53611931 & 0.49416880\\
 $1/3$ & 0.244 & 0.25342893 & 0.22409548\\
 $1/4$ & 0.158 & 0.74799673 & 0.73940730\\
 $1/4$ & 0.200 & 0.39547955 & 0.39700572\\
 $1/4$ & 0.241 & 0.09244687 & 0.11232334\\
 \hline
 \end{tabular}
\end{table}

The computed positivity of $q_0$ establishes stability against the radial
perturbations defined above, but it does not establish stability for an
arbitrary anisotropic material law. In particular, it does not remove
the equilibrium limitations associated with the $\eta=1$ family or justify
regions of tangential tension as a microscopic model of quark matter.
Nonradial stability, full characteristic causality and observational
viability require additional analysis. Moreover, the response laws are
constructed around each background; the sampled models are not by
themselves a continuous sequence with a single globally specified
material law. We therefore do not replace the eigenvalue calculation by
a mass-turning-point argument.

Finally, the exterior matching fixes the time normalization. Once a
physical radius is specified, the corresponding cyclic frequency is
\begin{equation}
 f=\frac{c}{2\pi R}\sqrt q.
 \label{rpnum:frequency-scale}
\end{equation}
The radius also fixes the density scale through $8\pi R^2\rho_s$ in
geometrized units. We therefore report dimensionless eigenvalues and do
not assign a frequency in kilohertz without specifying that scale.

To illustrate the physical scale of the results, we have constructed Table \ref{rpnum:dimensional-scales}, for which we have assumed $\rho_s = 240\, \mathrm{MeV/fm^{3}}$—a value that corresponds to $B = 60\,\mathrm{MeV/fm^{3}}$ when $\eta=1/3$ and $\rho_{s}=4B$. For other stiffness values, the same surface density is adopted as a reference, without implying that their equations of state follow the MIT model. In SI units, letting $\epsilon_s$ represent this energy density, we use $R=[s c^4/(8\pi G\epsilon_s)]^{1/2}$, $M=uRc^2/G$, and Eq.~\eqref{rpnum:frequency-scale}, with $s$ defined by Eq.~\eqref{eq:s5-surface-scale}. The values are calculated based on the eigenvalues provided in Table \ref{rpnum:fundamentals}. As the scale varies, $M$ and $R$ are proportional to $\rho_{s}^{-1/2}$, while $f_{0}$ is proportional to $\rho_{s}^{1/2}$.

\begin{table*}[t]
\centering
\caption{Dimensional scales of the twelve configurations for an illustrative surface density $\rho_s=240\,\mathrm{MeV/fm^3}$. The frequencies correspond to the fundamental mode and the two constitutive responses of the manuscript.}
\label{rpnum:dimensional-scales}
\begin{tabular}{cccccc}
\hline
$\eta$ & $u$ & $M/M_\odot$ & $R\ (\mathrm{km})$ & $f_0^{(\kappa=0)}\ (\mathrm{kHz})$ & $f_0^{(\kappa=1)}\ (\mathrm{kHz})$\\
\hline
$1$ & 0.100 & 0.574 & 8.478 & 13.665 & 13.465 \\
$1$ & 0.148 & 1.016 & 10.140 & 9.742 & 9.477 \\
$1$ & 0.196 & 1.511 & 11.382 & 7.109 & 6.767 \\
$1/2$ & 0.227 & 1.742 & 11.334 & 3.827 & 3.547 \\
$1/2$ & 0.243 & 1.885 & 11.454 & 3.289 & 2.980 \\
$1/2$ & 0.259 & 2.017 & 11.501 & 2.723 & 2.374 \\
$1/3$ & 0.181 & 1.273 & 10.384 & 4.257 & 4.138 \\
$1/3$ & 0.213 & 1.562 & 10.828 & 3.226 & 3.098 \\
$1/3$ & 0.244 & 1.822 & 11.029 & 2.178 & 2.048 \\
$1/4$ & 0.158 & 1.047 & 9.782 & 4.219 & 4.194 \\
$1/4$ & 0.200 & 1.417 & 10.464 & 2.867 & 2.873 \\
$1/4$ & 0.241 & 1.755 & 10.751 & 1.349 & 1.487 \\
\hline
\end{tabular}
\end{table*}

As an astrophysical reference, Gangopadhyay et al.~\cite{gangopadhyay2013strange} studied strange stars with masses on the order of $1$--$2\,M_\odot$ and compact radii; their model reaches a maximum mass of $2.0545\,M_\odot$ with a radius of $9.5\,\mathrm{km}$. The scales in Table~\ref{rpnum:dimensional-scales} allow us to place our configurations within the dimensional context of these models. This comparison is intended for guidance only: the equations of state and the assumptions regarding stresses differ, and it does not constitute a fit to those pulsars, nor does it determine the maximum mass for our families of stars. Furthermore, it does not provide an observational validation of the calculated frequencies.

The frequency scales in Table~\ref{rpnum:dimensional-scales} can also be placed in the context of previous radial-oscillation studies. V\'asquez Flores and Lugones~\cite{vazquezflores2010radial} found fundamental periods of approximately $0.1\,\mathrm{ms}$ for low-mass self-bound quark stars within a color-superconducting MIT bag framework, corresponding to frequencies of order $10\,\mathrm{kHz}$. Pretel et al.~\cite{pretel2024effects} further investigated how anisotropic pressure affects the masses, radii and radial oscillation spectra of interacting quark stars. These studies provide complementary context for our dimensional results; differences in the equations of state and anisotropic response prescriptions preclude a direct quantitative identification of their spectra with ours.

The following individual figures display all 96 modes, with displacement
and Lagrangian pressure perturbation shown separately. Each curve is divided
by its own maximum absolute value over the stored radial grid, so that the
figures compare radial shapes rather than amplitudes:
\begin{equation}
 \widehat\xi=\frac{\xi}{\max|\xi|},\qquad
 \widehat{\Delta_LP_r}=\frac{\Delta_LP_r}{\max|\Delta_LP_r|}.
 \label{rpnum:display-normalization}
\end{equation}
The common integration normalization is $y(0)=1$; the amplitude of a linear
eigenmode remains arbitrary. The absolute displacement vanishes at the
center, while the Lagrangian pressure perturbation vanishes at the surface.
The additional displacement nodes distinguish the three overtones from
the fundamental mode. Each mode profile is drawn as a solid colored line.

\begin{figure}[!htbp]
 \centering
 \includegraphics[width=0.90\columnwidth]{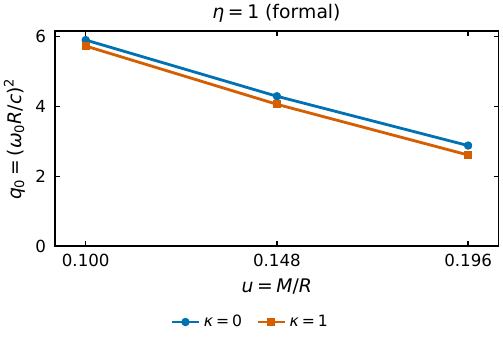}
 \caption{Fundamental radial eigenvalue $q_0$ for $\eta=1$ and
 $\kappa=0,1$. Solid lines join the three computed points as visual
 guides, without defining a continuously calculated sequence.}
 \label{rpnum:fundamental-eta1}
\end{figure}

\begin{figure}[!htbp]
 \centering
 \includegraphics[width=0.90\columnwidth]{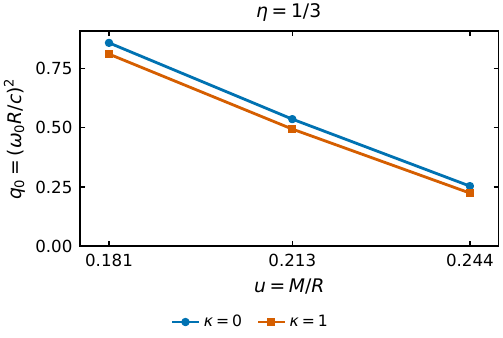}
 \caption{Fundamental radial eigenvalue $q_0$ for $\eta=1/3$ and
 $\kappa=0,1$. Solid lines join the three computed points as visual
 guides, without defining a continuously calculated sequence.}
 \label{rpnum:fundamental-eta3}
\end{figure}

\begin{figure}[!htbp]
 \centering
 \includegraphics[width=0.90\columnwidth]{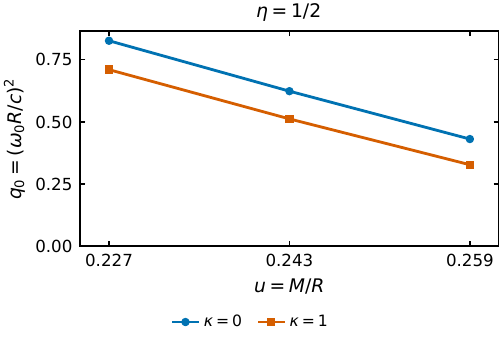}
 \caption{Fundamental radial eigenvalue $q_0$ for $\eta=1/2$ and
 $\kappa=0,1$. Solid lines join the three computed points as visual
 guides, without defining a continuously calculated sequence.}
 \label{rpnum:fundamental-eta2}
\end{figure}

\begin{figure}[!htbp]
 \centering
 \includegraphics[width=0.90\columnwidth]{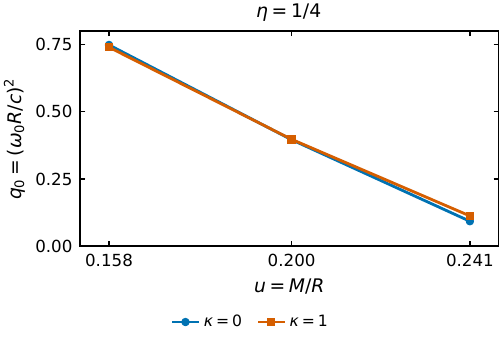}
 \caption{Fundamental radial eigenvalue $q_0$ for $\eta=1/4$ and
 $\kappa=0,1$. Solid lines join the three computed points as visual
 guides, without defining a continuously calculated sequence.}
 \label{rpnum:fundamental-eta4}
\end{figure}

\begin{figure*}[!t]
\centering
\subfloat[\label{rpnum:e1-k0-u1-displacement}]{%
  \includegraphics[width=0.48\textwidth]{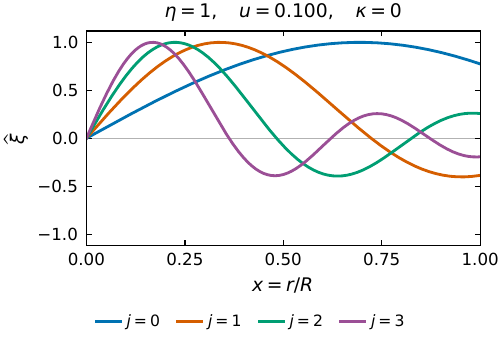}
}
\hfill
\subfloat[\label{rpnum:e1-k0-u1-pressure}]{%
  \includegraphics[width=0.48\textwidth]{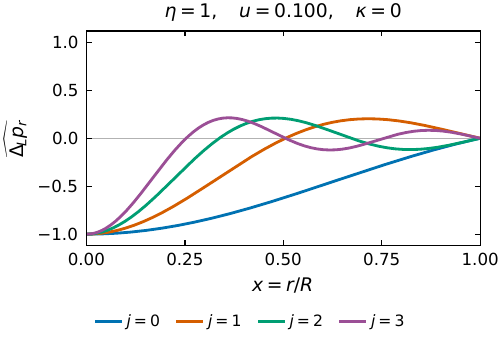}
}
\par\smallskip
\subfloat[\label{rpnum:e1-k0-u2-displacement}]{%
  \includegraphics[width=0.48\textwidth]{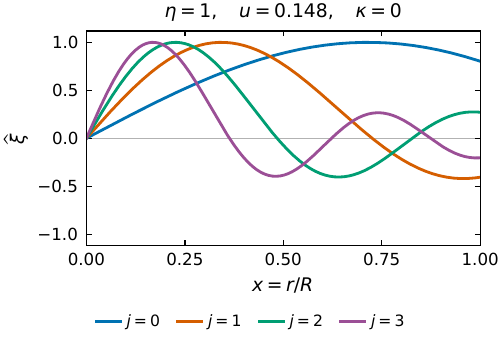}
}
\hfill
\subfloat[\label{rpnum:e1-k0-u2-pressure}]{%
  \includegraphics[width=0.48\textwidth]{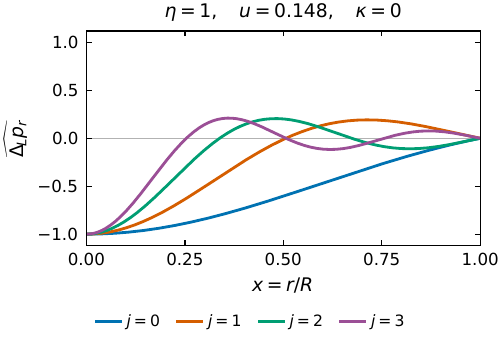}
}
\par\smallskip
\subfloat[\label{rpnum:e1-k0-u3-displacement}]{%
  \includegraphics[width=0.48\textwidth]{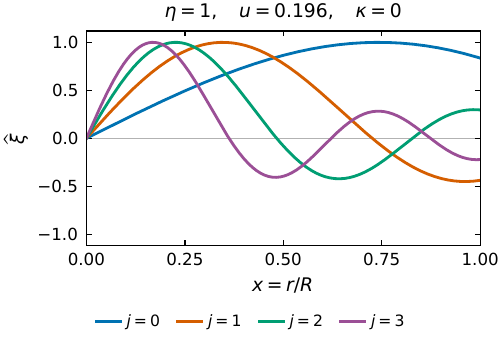}
}
\hfill
\subfloat[\label{rpnum:e1-k0-u3-pressure}]{%
  \includegraphics[width=0.48\textwidth]{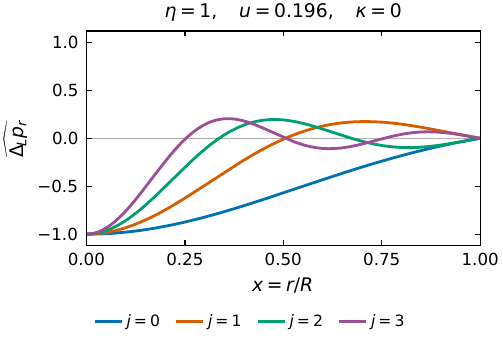}
}
\caption{Normalized radial modes for $\eta=1$ and $\kappa=0$. (a) $\widehat{\xi}$ and (b) $\widehat{\Delta_LP_r}$ for $u=0.100$; (c) $\widehat{\xi}$ and (d) $\widehat{\Delta_LP_r}$ for $u=0.148$; (e) $\widehat{\xi}$ and (f) $\widehat{\Delta_LP_r}$ for $u=0.196$. Each panel shows modes $j=0,1,2,3$ as solid colored curves, normalized separately as in Eq.~\eqref{rpnum:display-normalization}.}
\label{rpnum:e1-k0-modes}
\end{figure*}

\begin{figure*}[!t]
\centering
\subfloat[\label{rpnum:e1-k1-u1-displacement}]{%
  \includegraphics[width=0.48\textwidth]{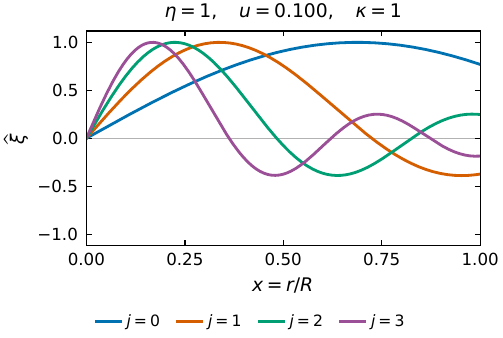}
}
\hfill
\subfloat[\label{rpnum:e1-k1-u1-pressure}]{%
  \includegraphics[width=0.48\textwidth]{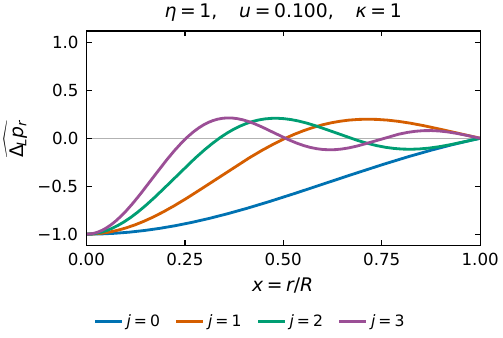}
}
\par\smallskip
\subfloat[\label{rpnum:e1-k1-u2-displacement}]{%
  \includegraphics[width=0.48\textwidth]{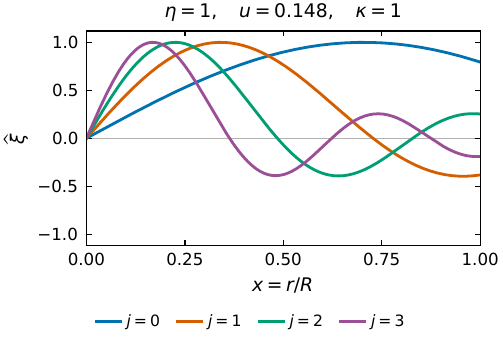}
}
\hfill
\subfloat[\label{rpnum:e1-k1-u2-pressure}]{%
  \includegraphics[width=0.48\textwidth]{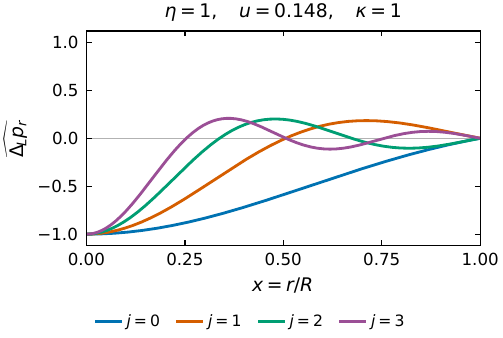}
}
\par\smallskip
\subfloat[\label{rpnum:e1-k1-u3-displacement}]{%
  \includegraphics[width=0.48\textwidth]{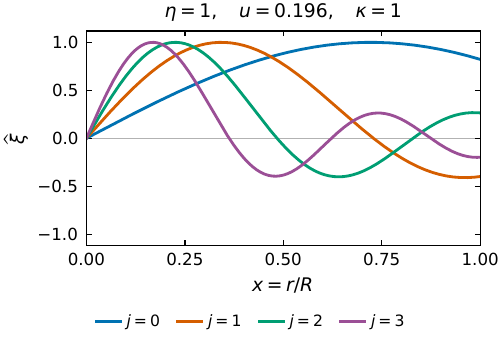}
}
\hfill
\subfloat[\label{rpnum:e1-k1-u3-pressure}]{%
  \includegraphics[width=0.48\textwidth]{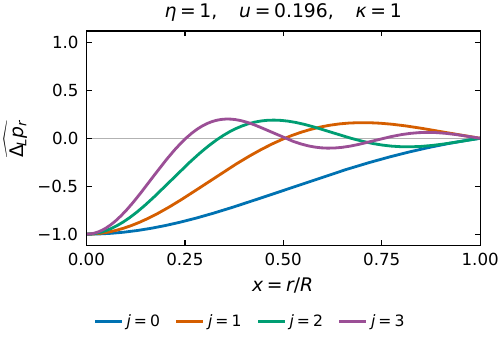}
}
\caption{Normalized radial modes for $\eta=1$ and $\kappa=1$. (a) $\widehat{\xi}$ and (b) $\widehat{\Delta_LP_r}$ for $u=0.100$; (c) $\widehat{\xi}$ and (d) $\widehat{\Delta_LP_r}$ for $u=0.148$; (e) $\widehat{\xi}$ and (f) $\widehat{\Delta_LP_r}$ for $u=0.196$. Each panel shows modes $j=0,1,2,3$ as solid colored curves, normalized separately as in Eq.~\eqref{rpnum:display-normalization}.}
\label{rpnum:e1-k1-modes}
\par\smallskip
\begin{minipage}{\textwidth}
\normalsize
For $\eta=1$, the individual profiles in
Figs.~\ref{rpnum:e1-k0-u1-displacement}--\ref{rpnum:e1-k1-u3-pressure}
retain the stiff family as a formal comparison. Both material responses
give positive fundamental eigenvalues at the three compactnesses shown.
The spectral result does not remove the physical reservations about this
equilibrium family discussed above.
\end{minipage}
\end{figure*}

\begin{figure*}[!t]
\centering
\subfloat[\label{rpnum:e2-k0-u1-displacement}]{%
  \includegraphics[width=0.48\textwidth]{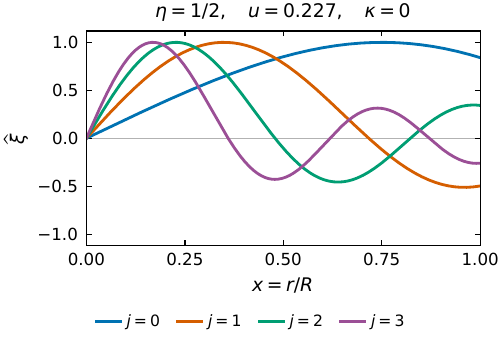}
}
\hfill
\subfloat[\label{rpnum:e2-k0-u1-pressure}]{%
  \includegraphics[width=0.48\textwidth]{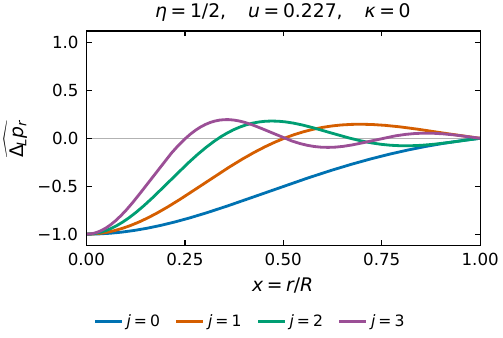}
}
\par\smallskip
\subfloat[\label{rpnum:e2-k0-u2-displacement}]{%
  \includegraphics[width=0.48\textwidth]{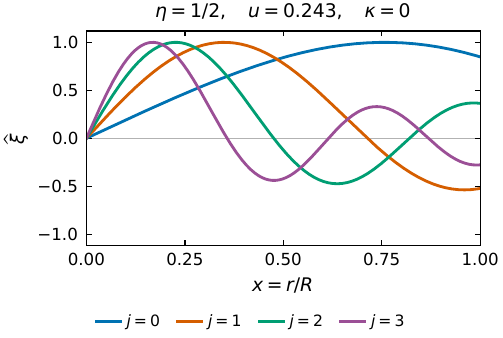}
}
\hfill
\subfloat[\label{rpnum:e2-k0-u2-pressure}]{%
  \includegraphics[width=0.48\textwidth]{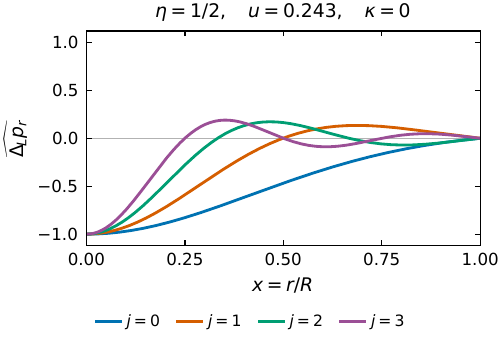}
}
\par\smallskip
\subfloat[\label{rpnum:e2-k0-u3-displacement}]{%
  \includegraphics[width=0.48\textwidth]{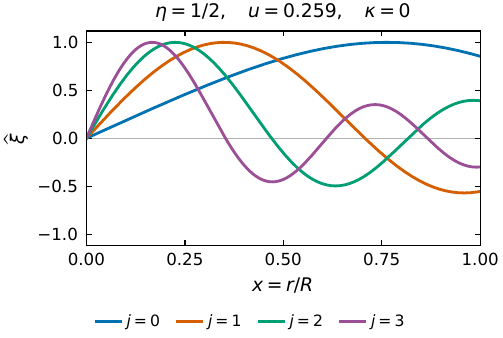}
}
\hfill
\subfloat[\label{rpnum:e2-k0-u3-pressure}]{%
  \includegraphics[width=0.48\textwidth]{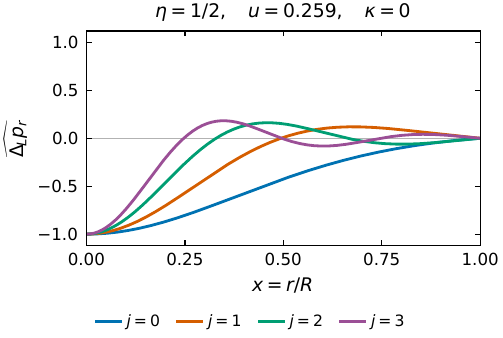}
}
\caption{Normalized radial modes for $\eta=1/2$ and $\kappa=0$. (a) $\widehat{\xi}$ and (b) $\widehat{\Delta_LP_r}$ for $u=0.227$; (c) $\widehat{\xi}$ and (d) $\widehat{\Delta_LP_r}$ for $u=0.243$; (e) $\widehat{\xi}$ and (f) $\widehat{\Delta_LP_r}$ for $u=0.259$. Each panel shows modes $j=0,1,2,3$ as solid colored curves, normalized separately as in Eq.~\eqref{rpnum:display-normalization}.}
\label{rpnum:e2-k0-modes}
\end{figure*}

\begin{figure*}[!t]
\centering
\subfloat[\label{rpnum:e2-k1-u1-displacement}]{%
  \includegraphics[width=0.48\textwidth]{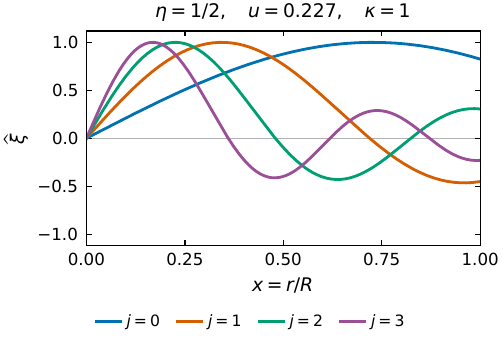}
}
\hfill
\subfloat[\label{rpnum:e2-k1-u1-pressure}]{%
  \includegraphics[width=0.48\textwidth]{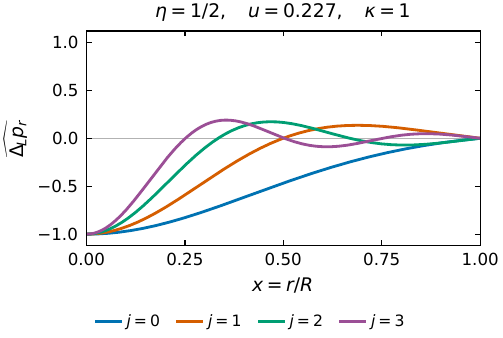}
}
\par\smallskip
\subfloat[\label{rpnum:e2-k1-u2-displacement}]{%
  \includegraphics[width=0.48\textwidth]{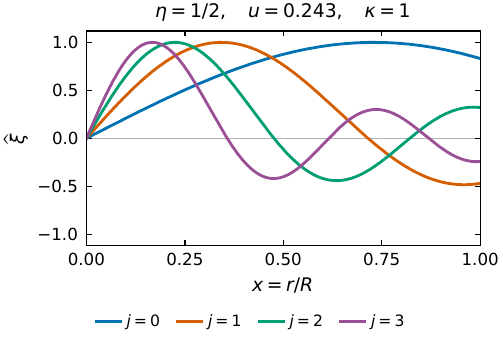}
}
\hfill
\subfloat[\label{rpnum:e2-k1-u2-pressure}]{%
  \includegraphics[width=0.48\textwidth]{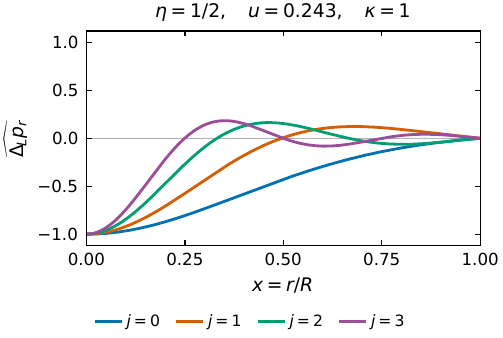}
}
\par\smallskip
\subfloat[\label{rpnum:e2-k1-u3-displacement}]{%
  \includegraphics[width=0.48\textwidth]{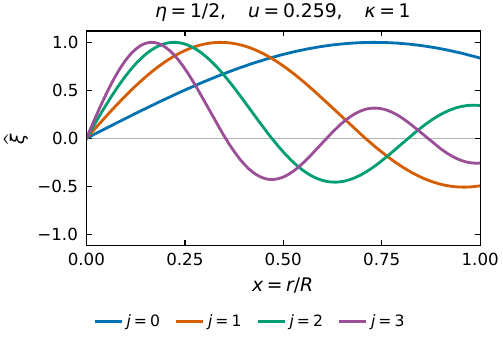}
}
\hfill
\subfloat[\label{rpnum:e2-k1-u3-pressure}]{%
  \includegraphics[width=0.48\textwidth]{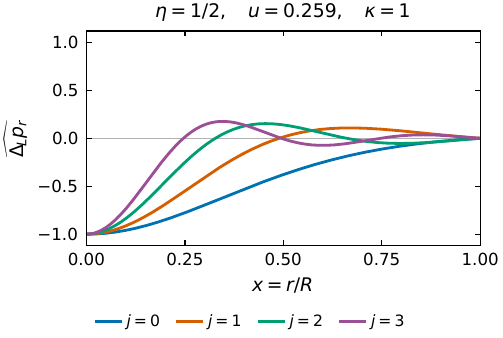}
}
\caption{Normalized radial modes for $\eta=1/2$ and $\kappa=1$. (a) $\widehat{\xi}$ and (b) $\widehat{\Delta_LP_r}$ for $u=0.227$; (c) $\widehat{\xi}$ and (d) $\widehat{\Delta_LP_r}$ for $u=0.243$; (e) $\widehat{\xi}$ and (f) $\widehat{\Delta_LP_r}$ for $u=0.259$. Each panel shows modes $j=0,1,2,3$ as solid colored curves, normalized separately as in Eq.~\eqref{rpnum:display-normalization}.}
\label{rpnum:e2-k1-modes}
\par\smallskip
\begin{minipage}{\textwidth}
\normalsize
For $\eta=1/2$, Figs.~\ref{rpnum:e2-k0-u1-displacement}--\ref{rpnum:e2-k1-u3-pressure}
show the displacement and pressure response at $u=0.227,0.243,0.259$.
The fundamental eigenvalue remains positive and decreases across these
sampled compactnesses for both closures. Changing the closure also changes
the eigenfunctions, even though the equilibrium is held fixed.
\end{minipage}
\end{figure*}

\begin{figure*}[!t]
\centering
\subfloat[\label{rpnum:e3-k0-u1-displacement}]{%
  \includegraphics[width=0.48\textwidth]{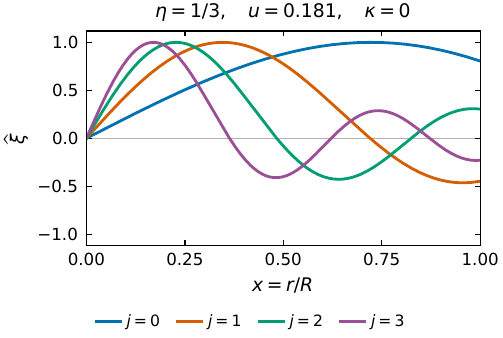}
}
\hfill
\subfloat[\label{rpnum:e3-k0-u1-pressure}]{%
  \includegraphics[width=0.48\textwidth]{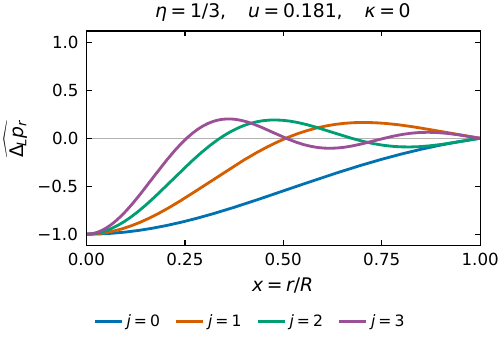}
}
\par\smallskip
\subfloat[\label{rpnum:e3-k0-u2-displacement}]{%
  \includegraphics[width=0.48\textwidth]{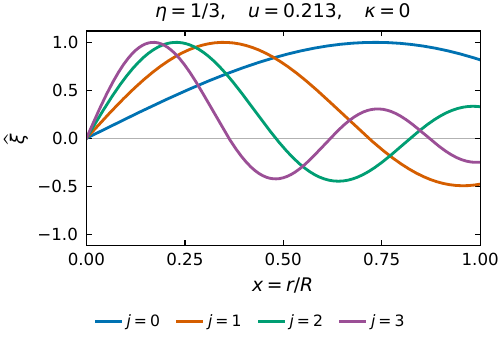}
}
\hfill
\subfloat[\label{rpnum:e3-k0-u2-pressure}]{%
  \includegraphics[width=0.48\textwidth]{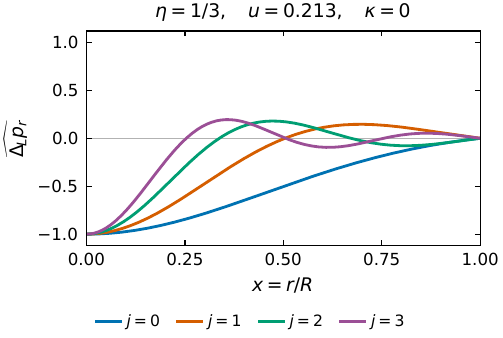}
}
\par\smallskip
\subfloat[\label{rpnum:e3-k0-u3-displacement}]{%
  \includegraphics[width=0.48\textwidth]{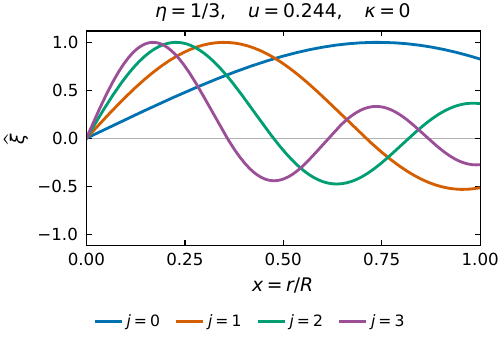}
}
\hfill
\subfloat[\label{rpnum:e3-k0-u3-pressure}]{%
  \includegraphics[width=0.48\textwidth]{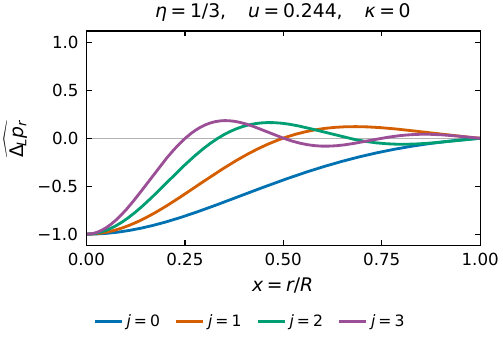}
}
\caption{Normalized radial modes for $\eta=1/3$ and $\kappa=0$. (a) $\widehat{\xi}$ and (b) $\widehat{\Delta_LP_r}$ for $u=0.181$; (c) $\widehat{\xi}$ and (d) $\widehat{\Delta_LP_r}$ for $u=0.213$; (e) $\widehat{\xi}$ and (f) $\widehat{\Delta_LP_r}$ for $u=0.244$. Each panel shows modes $j=0,1,2,3$ as solid colored curves, normalized separately as in Eq.~\eqref{rpnum:display-normalization}.}
\label{rpnum:e3-k0-modes}
\end{figure*}

\begin{figure*}[!t]
\centering
\subfloat[\label{rpnum:e3-k1-u1-displacement}]{%
  \includegraphics[width=0.48\textwidth]{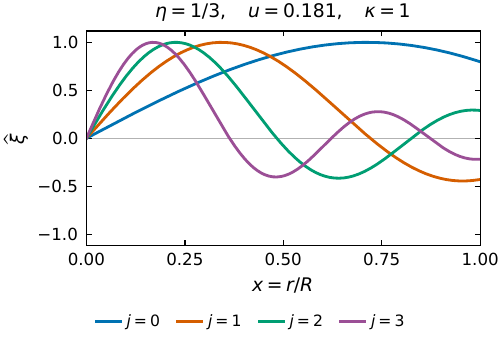}
}
\hfill
\subfloat[\label{rpnum:e3-k1-u1-pressure}]{%
  \includegraphics[width=0.48\textwidth]{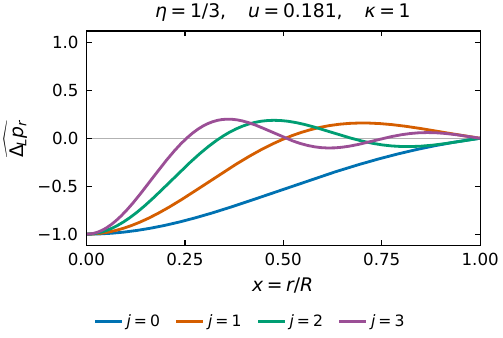}
}
\par\smallskip
\subfloat[\label{rpnum:e3-k1-u2-displacement}]{%
  \includegraphics[width=0.48\textwidth]{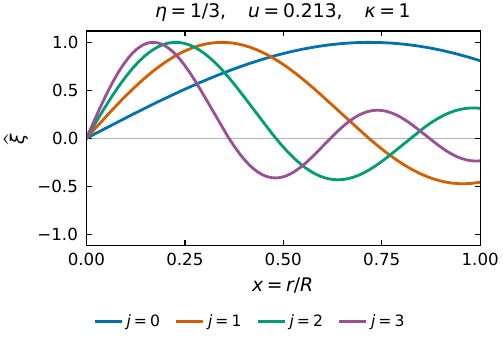}
}
\hfill
\subfloat[\label{rpnum:e3-k1-u2-pressure}]{%
  \includegraphics[width=0.48\textwidth]{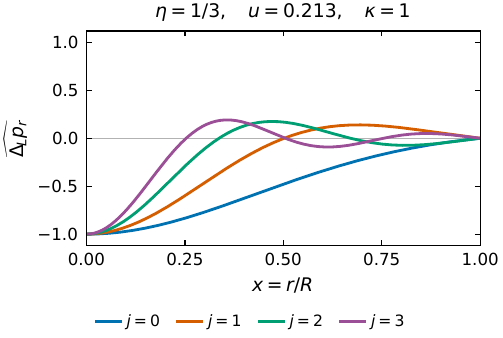}
}
\par\smallskip
\subfloat[\label{rpnum:e3-k1-u3-displacement}]{%
  \includegraphics[width=0.48\textwidth]{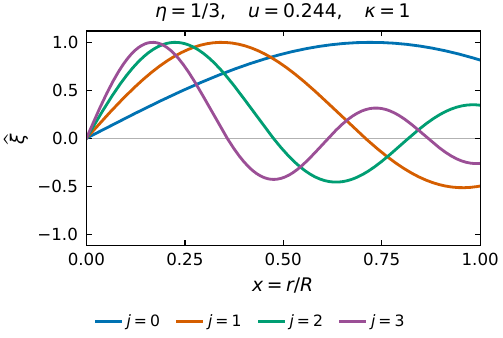}
}
\hfill
\subfloat[\label{rpnum:e3-k1-u3-pressure}]{%
  \includegraphics[width=0.48\textwidth]{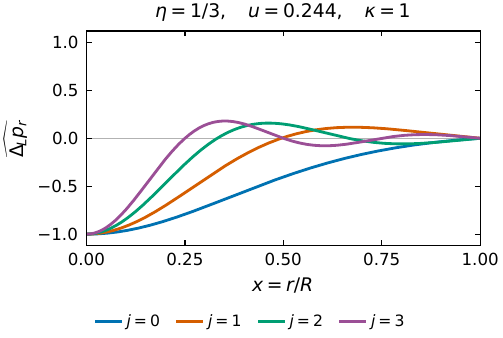}
}
\caption{Normalized radial modes for $\eta=1/3$ and $\kappa=1$. (a) $\widehat{\xi}$ and (b) $\widehat{\Delta_LP_r}$ for $u=0.181$; (c) $\widehat{\xi}$ and (d) $\widehat{\Delta_LP_r}$ for $u=0.213$; (e) $\widehat{\xi}$ and (f) $\widehat{\Delta_LP_r}$ for $u=0.244$. Each panel shows modes $j=0,1,2,3$ as solid colored curves, normalized separately as in Eq.~\eqref{rpnum:display-normalization}.}
\label{rpnum:e3-k1-modes}
\par\smallskip
\begin{minipage}{\textwidth}
\normalsize
For $\eta=1/3$, Figs.~\ref{rpnum:e3-k0-u1-displacement}--\ref{rpnum:e3-k1-u3-pressure}
show all four modes at $u=0.181,0.213,0.244$. The displacement profiles
exhibit the expected progression from the nodeless fundamental mode to
three nodes in the third overtone. The surface condition applies to the
Lagrangian pressure perturbation in each of the accompanying profiles.
\end{minipage}
\end{figure*}

\begin{figure*}[!t]
\centering
\subfloat[\label{rpnum:e4-k0-u1-displacement}]{%
  \includegraphics[width=0.48\textwidth]{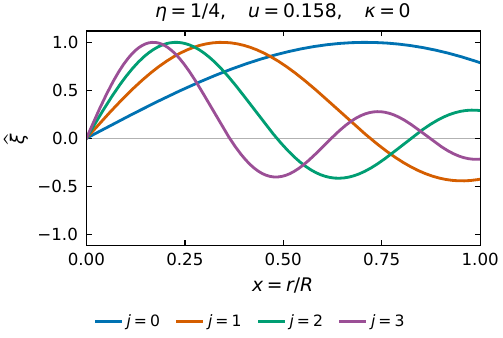}
}
\hfill
\subfloat[\label{rpnum:e4-k0-u1-pressure}]{%
  \includegraphics[width=0.48\textwidth]{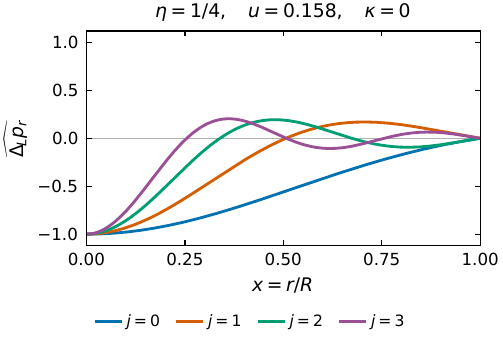}
}
\par\smallskip
\subfloat[\label{rpnum:e4-k0-u2-displacement}]{%
  \includegraphics[width=0.48\textwidth]{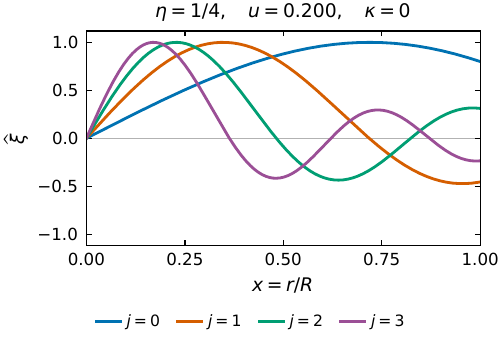}
}
\hfill
\subfloat[\label{rpnum:e4-k0-u2-pressure}]{%
  \includegraphics[width=0.48\textwidth]{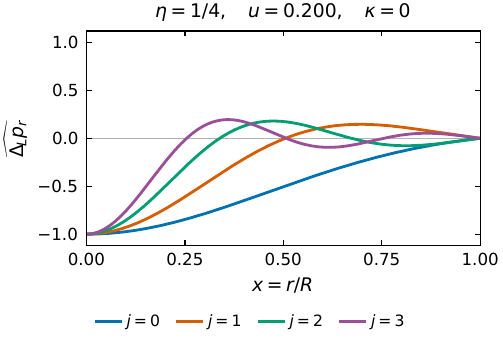}
}
\par\smallskip
\subfloat[\label{rpnum:e4-k0-u3-displacement}]{%
  \includegraphics[width=0.48\textwidth]{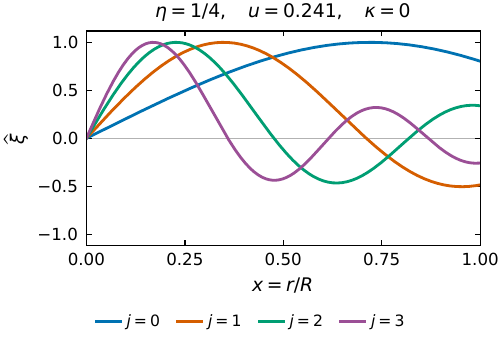}
}
\hfill
\subfloat[\label{rpnum:e4-k0-u3-pressure}]{%
  \includegraphics[width=0.48\textwidth]{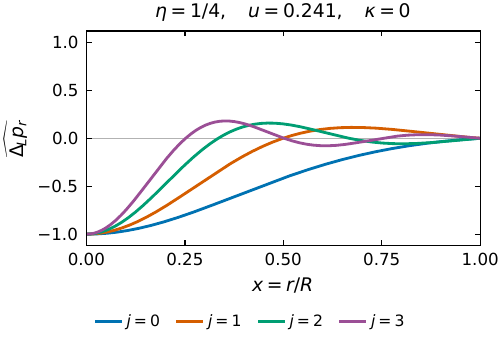}
}
\caption{Normalized radial modes for $\eta=1/4$ and $\kappa=0$. (a) $\widehat{\xi}$ and (b) $\widehat{\Delta_LP_r}$ for $u=0.158$; (c) $\widehat{\xi}$ and (d) $\widehat{\Delta_LP_r}$ for $u=0.200$; (e) $\widehat{\xi}$ and (f) $\widehat{\Delta_LP_r}$ for $u=0.241$. Each panel shows modes $j=0,1,2,3$ as solid colored curves, normalized separately as in Eq.~\eqref{rpnum:display-normalization}.}
\label{rpnum:e4-k0-modes}
\end{figure*}

\begin{figure*}[!t]
\centering
\subfloat[\label{rpnum:e4-k1-u1-displacement}]{%
  \includegraphics[width=0.48\textwidth]{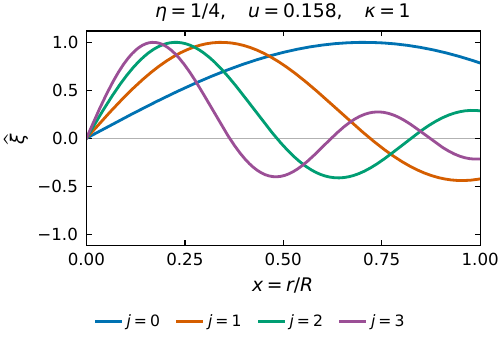}
}
\hfill
\subfloat[\label{rpnum:e4-k1-u1-pressure}]{%
  \includegraphics[width=0.48\textwidth]{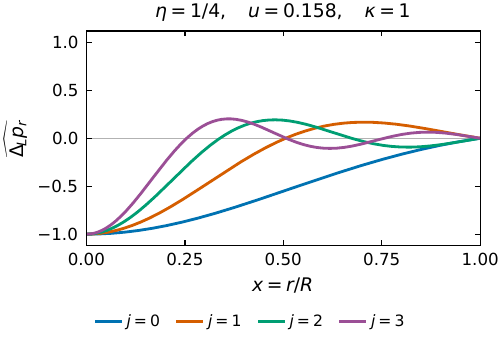}
}
\par\smallskip
\subfloat[\label{rpnum:e4-k1-u2-displacement}]{%
  \includegraphics[width=0.48\textwidth]{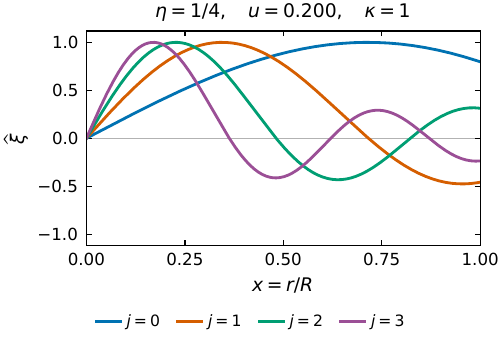}
}
\hfill
\subfloat[\label{rpnum:e4-k1-u2-pressure}]{%
  \includegraphics[width=0.48\textwidth]{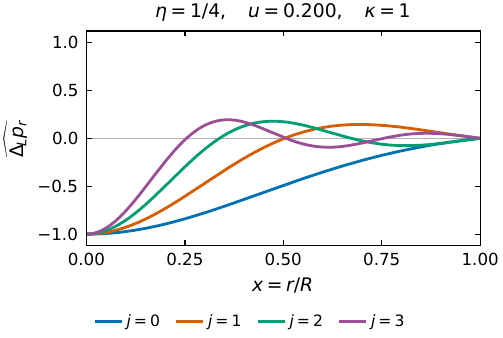}
}
\par\smallskip
\subfloat[\label{rpnum:e4-k1-u3-displacement}]{%
  \includegraphics[width=0.48\textwidth]{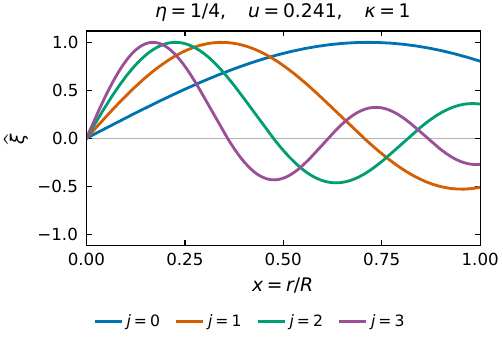}
}
\hfill
\subfloat[\label{rpnum:e4-k1-u3-pressure}]{%
  \includegraphics[width=0.48\textwidth]{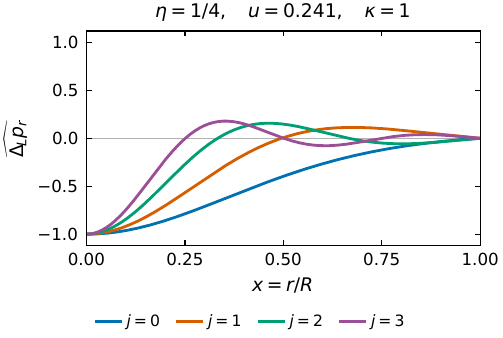}
}
\caption{Normalized radial modes for $\eta=1/4$ and $\kappa=1$. (a) $\widehat{\xi}$ and (b) $\widehat{\Delta_LP_r}$ for $u=0.158$; (c) $\widehat{\xi}$ and (d) $\widehat{\Delta_LP_r}$ for $u=0.200$; (e) $\widehat{\xi}$ and (f) $\widehat{\Delta_LP_r}$ for $u=0.241$. Each panel shows modes $j=0,1,2,3$ as solid colored curves, normalized separately as in Eq.~\eqref{rpnum:display-normalization}.}
\label{rpnum:e4-k1-modes}
\par\smallskip
\begin{minipage}{\textwidth}
\normalsize
For $\eta=1/4$, Figs.~\ref{rpnum:e4-k0-u1-displacement}--\ref{rpnum:e4-k1-u3-pressure}
complete the mode comparison at $u=0.158,0.200,0.241$. This family contains
the smallest computed fundamental eigenvalue. Both closures still give
$q_0>0$, while their ordering changes between the lowest and the two
higher sampled compactnesses.
\end{minipage}
\end{figure*}

\clearpage
\onecolumn
\twocolumn

\section{Discussion and conclusions}
\label{section-conclusions}

The present construction supplements EFEs with two explicit assumptions: the Tolman IV temporal potential and a generalized linear EoS for the complete energy density and radial pressure. The EoS determines the radial deformation, and the remaining field equation determines the tangential stress. This separates the geometric choice defining the equilibrium family from the material-response prescriptions needed to determine its radial spectrum. 

The field equations involve the metric potentials and the thermodynamic variables $\rho$, $P_r$ and $P_{\perp}$, and therefore require additional relations to form a closed system. In our construction, the prescribed temporal potential and the radial relation $P_r=\eta(\rho-\rho_s)$ determine the complete radial metric potential, after central regularity and Schwarzschild matching are imposed. The tangential pressure and the anisotropy then follow from the remaining field equation. Thus, the radial EoS does not independently prescribe the tangential sector, and minimal geometric deformation provides a consistent representation of the additional source rather than a microscopic explanation of its stresses. The resulting equilibrium solution admits a closed rational representation in terms of finite polynomial sums for $\eta=1,1/2,1/3$ and $1/4$.

The generalized linear parametrization separates the stiffness of the radial response, controlled by $\eta$, from the surface-density scale set by $\rho_s$. This separation provides a framework for investigating changes in stellar structure and oscillation spectra, but the parameters of the matched family cannot all be varied independently. In particular, Eq.~\eqref{eq:s5-surface-scale} fixes $s=8\pi R^2\rho_s$ once $u$ and $\eta$ are chosen. At fixed $u$ and $\eta$, the dimensional mass and radius therefore scale as $\rho_s^{-1/2}$, while the oscillation frequencies scale as $\rho_s^{1/2}$. Comparisons at fixed physical EoS must respect this scaling; plots with $R=1$ do not represent stars with a common dimensional radius. 

For $\eta=1/3$ and $\rho_s=4B$, the radial EoS reduces to the simplest MIT bag relation. This identifies a thermodynamic limit of the radial sector, without making the complete anisotropic configuration equivalent to an isotropic MIT bag star. The finite interior surface density follows from $P_r(R)=0$ and represents the self-bound boundary of the model. The displayed equilibria are regular and satisfy the reported energy inequalities, but their tangential stresses distinguish the families. Tangential tension occurs near the surface in the sampled $\eta=1$ and $\eta=1/2$ configurations and in the two less compact $\eta=1/3$ examples. By contrast, $P_{\perp}$ remains positive in the sampled $\eta=1/4$ models and in the $\eta=1/3$, $u=0.244$ configuration. The $\eta=1$ family also fails the adopted subluminal tangential-profile diagnostic and is retained only as a formal comparison. Neither the energy inequalities nor radial-mode stability removes these limitations.

The distinction between equilibrium closure and dynamical closure is essential. A static anisotropy profile does not determine its response to a fluid displacement, so two material prescriptions were specified: conservation of the anisotropy carried by each layer for $\kappa=0$, and conservation of its ratio to $\rho+P_r$ for $\kappa=1$. Both reproduce the same background while defining different perturbation problems. With metric perturbations included, central regularity and the moving-surface condition $\Delta_LP_r(R)=0$ yield the radial spectrum. The nonzero surface density causes no singularity in the relevant adiabatic coefficient, since $\Gamma_rP_r=\eta(\rho+P_r)$ remains finite even though $\Gamma_r$ diverges as the radial pressure vanishes.

The fundamental mode and the first three overtones were computed for twelve backgrounds and both material prescriptions, giving 96 eigenvalues. All 24 background--closure combinations have a positive fundamental eigenvalue and are therefore linearly stable against the specified radial perturbations. Within each sampled stiffness family, $q_0$ decreases with increasing compactness for both closures. The smallest reported value is $q_0=0.09244687$, obtained for $\eta=1/4$, $u=0.241$ and $\kappa=0$. Nevertheless, changing the anisotropic response does not produce a universal shift: $\kappa=1$ lowers the fundamental eigenvalue in the sampled $\eta=1,1/2,1/3$ families, whereas the ordering reverses at the two larger compactnesses of the $\eta=1/4$ family. At $\eta=1/4$ and $u=0.241$, this change of response increases the fundamental frequency by approximately $10.2\%$ at fixed background, despite leaving the equilibrium geometry unchanged. The equilibrium geometry alone is therefore insufficient to determine the radial spectrum.

The importance of specifying the anisotropic response and the surface conditions is consistent with the analysis of Arba\~nil and Malheiro~\cite{arbanil2016anisotropic}, who showed that the relation between maximum mass and the onset of radial instability in anisotropic strange stars depends on the equilibrium sequence and its surface anisotropy. Our discrete samples, with response laws defined around each background, do not establish a continuous sequence governed by a single global material prescription. Consequently, the present results neither identify a maximum stable mass nor locate a critical compactness where the fundamental mode vanishes. They also do not establish nonradial stability, full characteristic causality or agreement with observational constraints.

Finally, the main outcome is an analytically tractable family of self-bound anisotropic equilibria together with a radial stability analysis that explicitly separates the assumptions defining the background from those defining its material response. Extending the calculation to continuous sequences at fixed surface density and with a globally specified anisotropic constitutive law would allow stability boundaries and mass-radius relations to be assessed consistently. A microscopic account of the tangential stresses, particularly in regions of tension, and an analysis of nonradial modes and tidal response are further steps needed to evaluate the astrophysical interpretation of these configurations.

\end{document}